\documentclass[10pt,aps,prl,floatfix,showpacs,twocolumn,longbibliography,superscriptaddress]{revtex4-1}
\usepackage{ulem}
\usepackage{dcolumn}
\usepackage{bm}
\usepackage{color}
\usepackage{amssymb}
\usepackage{amsmath}
\usepackage{graphicx}
\usepackage{amsfonts}
\usepackage{slashed}
\usepackage{pstricks}
\usepackage{hyperref}
\usepackage{array}
\usepackage{multirow}
\usepackage{array}
\allowdisplaybreaks
\usepackage{dcolumn}
\usepackage{epsf}
\usepackage[nice]{nicefrac}
\usepackage{tikz}
\usepackage{MnSymbol}

\newcommand{\Lag}{\mathcal{L}}
\newcommand{\calF}{\mathcal{F}}
\newcommand{\calR}{\mathcal{R}}
\begin{document}

\title{Nuclear responses with neural-network quantum states.}
\author{Elad Parnes}
\email{elad.parnes@mail.huji.ac.il}
\affiliation{Racah Institute of Physics, The Hebrew University, Jerusalem 91904, Israel}

\author{Nir Barnea}
\email{nir.barnea1@mail.huji.ac.il}
\affiliation{Racah Institute of Physics, The Hebrew University, Jerusalem 91904, Israel}

\author{Giuseppe Carleo}
\email{giuseppe.carleo@epfl.ch}
\affiliation{Institute of Physics, École Polytechnique Fédérale de Lausanne (EPFL), CH-1015 Lausanne, Switzerland}

\author{Alessandro Lovato}
\email{lovato@anl.gov}
\affiliation{Physics Division, Argonne National Laboratory, Argonne, Illinois 60439, USA}
\affiliation{Computational Science (CPS) Division, Argonne National Laboratory, Argonne, Illinois 60439, USA}
\affiliation{INFN-TIFPA Trento Institute for Fundamental Physics and Applications, Trento, Italy}

\author{Noemi Rocco}
\email{nrocco@fnal.gov}
\affiliation{Theoretical Physics Department Fermi National Accelerator Laboratory P.O. Box 500 Batavia Illinois 60510 USA}

\author{Xilin Zhang}
\email{zhangx@frib.msu.edu}
\affiliation{Facility for Rare Isotope Beams, Michigan State University, Michigan 48824, USA}

\date{\today}

\begin{abstract}
We introduce a variational Monte Carlo framework that combines neural-network quantum states with the Lorentz integral transform technique to compute the dynamical properties of self-bound quantum many-body systems in continuous Hilbert spaces. While broadly applicable to various quantum systems, including atoms and molecules, in this initial application we focus on the photoabsorption cross section of light nuclei, where benchmarks against numerically exact techniques are available. Our accurate theoretical predictions are complemented by robust uncertainty quantification, enabling meaningful comparisons with experiments. We demonstrate that a simple nuclear Hamiltonian, based on a leading-order pionless effective field theory expansion and known to accurately reproduce the ground-state energies of nuclei with $A\leq 20$ nucleons also provides a reliable description of the photoabsorption cross section.
\end{abstract}

\maketitle

\textit{Introduction.}---
The past decades have witnessed a dramatic expansion in the scope of nuclear ab initio methods~\cite{Hergert:2020bxy}. Ground-state properties and electroweak transitions of medium-mass and heavy nuclei can now be described as emergent phenomena arising from the individual interactions among neutrons and protons, and are supplemented by estimates of theoretical uncertainties~\cite{Hagen:2016uwj,Morris:2017vxi,Gysbers:2019uyb,Hu:2021trw}. Notably, these methods have also advanced in predicting dynamical properties of such systems, particularly within the linear response regime, which is well suited for modeling interactions between atomic nuclei and electroweak probes~\cite{Bacca:2014tla}.

These calculations are crucial for interpreting electron-scattering experiments, including determining whether explicit QCD effects are necessary to explain the measured response functions~\cite{Lovato:2013cua,Cloet:2015tha,Lovato:2016gkq}. Moreover, they are critical for fully exploiting current and next-generation accelerator-based neutrino oscillation experiments~\cite{Acharya:2024xah,Lovato:2020kba}, which use atomic nuclei in their detectors to enhance event rates. Finally, they are also essential for understanding aspects of neutron-star cooling, which is primarily driven by neutrino emission~\cite{Yakovlev:2004iq,Shen:2012sa,Sobczyk:2024hdl}.

Continuum quantum Monte Carlo (QMC) methods~\cite{Carlson:2014vla} can incorporate high-resolution nuclear forces to describe lepton-nucleus scattering at relatively high momentum transfers~\cite{Lovato:2016gkq,Lovato:2020kba,Nikolakopoulos:2023zse}. However, their exponentially increasing computational cost and the onset of the fermion-sign problem limit their applicability to relatively light nuclei --- see Ref.~\cite{Gnech:2024qru} for exploratory applications to the isoscalar density response of $^{16}$O. Additionally, the use of the Laplace transform hinders the resolution of rich features in the low-energy spectrum, characterized by discrete transitions and collective modes. Machine-learning approaches that have recently been devised to invert the Laplace transform~\cite{Raghavan:2020bze,Raghavan:2023pav} mitigate but do not completely resolve this shortcoming.

Coupled-cluster theory~\cite{Hagen:2013nca} has been successfully employed to compute response functions relevant for both inclusive electron- and neutrino-nucleus scattering cross sections for nuclei as large as $^{16}$O and $^{40}$Ca~\cite{Sobczyk:2021dwm,Sobczyk:2023sxh,Acharya:2024xah}. Lower energy observables, such as dipole strength functions~\cite{Miorelli:2016qbk,Birkhan:2016qkr,Fearick:2023lyz} and polarizability, have also been computed, including for open-shell nuclei~\cite{Bonaiti:2021kkp,Bonaiti:2024fft}. The symmetry-adapted no-core shell model has recently been applied to elucidate the dynamical properties of atomic nuclei~\cite{Burrows:2023ugy}. However, both methods rely on single-particle basis expansions and thus struggle to incorporate high-resolution interactions.

Following their seminal applications to systems of interacting spins~\cite{Carleo:2017} and quantum chemistry problems~\cite{Hermann:2020xqs,Pfau:2020}, artificial neural networks have been increasingly used to compactly and efficiently represent nuclear quantum many-body wave functions~\cite{Keeble:2019bkv,Adams:2020aax,Lovato:2022tjh,Yang:2022esu}. The adoption of these neural quantum states (NQS) has enabled the extension of QMC to nuclei with up to $A=20$ nucleons~\cite{Gnech24}, while recent studies~\cite{Fore:2024exa} have performed infinite nuclear matter calculations with up to $A=42$ nucleons.

In this work, we extend the application of NQS to study the dynamical properties of atomic nuclei in the linear response regime. Specifically, we combine the Lorentz integral transform (LIT) technique~\cite{Efros:2007nq} with the method introduced in Ref.~\cite{Hendry:2019} to compute the spectral functions of interacting spin systems, which was, in turn, inspired by earlier density matrix renormalization group applications~\cite{Kuhner:1999}. Beyond providing accurate estimates of the LIT, our approach allows for precise upper bounds on the theoretical uncertainty of our calculations. We focus on the dipole response of atomic nuclei, specifically their total photoabsorption cross sections, which provide critical insight into nuclear dynamics~\cite{Bacca:2014tla}. More broadly, our method is applicable to the computation of response functions for any self-bound quantum many-body system described by NQS in continuum Hilbert spaces and can be readily applied, for instance, to atoms and molecules.

\textit{Method.}--- 
In the linear response regime, the interactions of atomic nuclei with electroweak probes are modeled by the nuclear response functions
\begin{equation} 
  \calR(\omega)=\sumint_f |\langle\Psi_f | \hat O | \Psi_0 \rangle |^2 \delta{(E_f-E_0-\omega)} \,,
\end{equation}
which encapsulate the dynamical part of the reaction.
In the above equation, \( \hat O\) denotes the excitation operator, while \( | \Psi_0 \rangle \) and \( | \Psi_f \rangle \) represent the initial and final states of the system, with energies \(E_0\) and \(E_f\), respectively. The final state can be either bound (belonging to the discrete part of the spectrum) or unbound (in the continuum). A direct calculation of the response function requires evaluating all transition amplitudes. Integral transform techniques, such as the Laplace transform~\cite{Carlson:1992ga,Carlson:1994zz} and the LIT~\cite{Efros:2007nq}, circumvent the difficulties associated with the direct calculation of continuum final states. The LIT is defined as 
\begin{equation} \label{LIT}
  {\cal L}(\omega_0,\Gamma)=\int_{\omega_{th}}^\infty d\omega \frac{\calR(\omega)}{(\omega-\omega_0)^2+\Gamma^2} \;,
\end{equation}
where \(\omega_{th}\) is the threshold energy and \(\Gamma>0\) is the Lorentzian width, serving as a resolution parameter. Once \({\cal L}(\omega_0,\Gamma)\) is computed by means of few- or many-body techniques, inverting Eq. (\ref{LIT}) yields the response function. Utilizing the completeness of the final states, one can show that $
{\cal L}(\omega_0,\Gamma)=\langle \Psi_L | \Psi_L \rangle$, where \(|\Psi_L\rangle\) is the solution of the inhomogeneous  Schr\"odinger equation
\begin{equation}\label{psitilde}
 ( \hat H - z) |\Psi_L\rangle = \hat O | \Psi_0 \rangle \;,
\end{equation}
where \(z = E_0 + \omega_0 + i\Gamma\). The response function can, in principle, be directly recovered by taking \(\Gamma \to 0\). However, in that case, \(\Psi_L\) becomes unbound, making its calculation for a generic many-body system unfeasible. 
To avoid this difficulty, we keep \(\Gamma\) finite. 
%--- specifically, we choose either \(\Gamma = 5\) MeV 
%or \(\Gamma = 10\) MeV --- 
so that, due to the localized nature of the source term, the behavior of \(\Psi_L\) resembles that of bound-state wave functions.

We parameterize both \(\Psi_0(X) \equiv \langle X | \Psi_0 \rangle\) and \(\Psi_L(X) \equiv \langle X | \Psi_L \rangle\) using NQS. Here, $X=\{x_1\dots x_A\}$ denotes the set of single-particle coordinates $x_i = \{\mathbf{r}_i, s^z_i, t^z_i\}$, which describe the spatial positions and the z-projection of the spin-isospin degrees of freedom of the $A$ nucleons. Our NQS of choice is the neural Pfaffian introduced in Ref.~\cite{Kim:2023fwy} for ultracold Fermi gases and recently generalized to nuclear systems~\cite{Fore:2024exa}. As for the Hamiltonian, we take the model ``o'' of~\cite{Schiavilla:2021dun}, which is derived from a leading-order pionless effective field theory expansion and reproduces well the ground state energy of $A\leq 20$ nuclei~\cite{Gnech:2024qru}.  

Following Ref.~\cite{Hendry:2019}, we solve the LIT equation by maximizing the quantum fidelity between the states \( |\Psi\rangle \equiv (\hat H-z)|\Psi_L\rangle \) and \( |\Phi\rangle \equiv \hat O | \Psi_0 \rangle \), which is defined as~\cite{Sinibaldi:2023yoq}
\begin{equation}
\mathcal{F}(\Psi, \Phi) = \frac{\langle \Psi | \Phi \rangle \langle \Phi | \Psi \rangle }{ \langle \Psi | \Psi \rangle \langle \Phi | \Phi \rangle}.
\end{equation}
The optimal parameters \( \theta \) of the NQS are found using the stochastic reconfiguration method~\cite{Sorella:2005}, which accounts for the curved geometry of the Hilbert space. The parameter update is given by \( \delta \theta = \eta S^{-1} \nabla_{\theta}\mathcal{F} \), where \( \eta \) is the learning rate, \( \nabla_{\theta}\mathcal{F} \) is the gradient of the fidelity, and
\begin{equation}
S_{ij} = \frac{\langle \partial_{\theta_i} \Psi | \partial_{\theta_j}\Psi \rangle}{\langle \Psi | \Psi \rangle } - \frac{\langle \partial_{\theta_i} \Psi | \Psi \rangle}{\langle \Psi | \Psi \rangle } \frac{\langle \Psi |  \partial_{\theta_j}\Psi \rangle}{\langle \Psi | \Psi \rangle }
\label{eq:fisher}
\end{equation}
is the quantum Fisher information matrix.

The fidelity, its gradient, and the quantum Fisher information matrix are estimated using Monte Carlo techniques to sample the probability distributions \( \pi_\Psi(X) = |\Psi(X)|^2 / \langle \Psi | \Psi \rangle \) and \( \pi_\Phi(X) = |\Phi(X)|^2 / \langle \Phi | \Phi \rangle \). To reduce statistical noise, we employ the control-variate estimators recently proposed in Ref.~\cite{Gravina:2024elh}. Additionally, we reduce the computational cost of sampling from \( \pi_\Psi(X) \) by reweighting samples drawn from \( \pi_\Phi(X) \). Since the latter probability distribution does not depend on \( z \), we can reuse the same samples for all values of \( \omega_0 \). The full expressions for the above estimators are provided in the Supplemental Material. 

Maximizing the fidelity yields a state that is as parallel as possible, within the variational space, to the target state, but it does not constrain its phase and norm. In other words, it gives $|\bar{\Psi}\rangle \approx \mathcal{N} |\Psi\rangle $, where $|\Psi\rangle$ is the solution we seek~\cite{Hendry:2019}. This complex constant is given by $\mathcal{N} = \langle \Phi | \bar{\Psi}\rangle / \langle \Phi | \Phi \rangle$, and the latter ratio is estimated stochastically by sampling from $\pi_{\Phi}(X)$. 

Computing the LIT as the norm of \(\Psi_L\) is computationally challenging due to the slowly decaying and oscillatory tails of this wave function. To address this difficulty, we rewrite the norm as  
\begin{equation}
\langle\Psi_{L}|\Psi_{L}\rangle=\langle\Phi|\frac{1}{\hat{H}-z^{*}}\frac{1}{\hat{H}-z}|\Phi\rangle\nonumber=\frac{1}{\Gamma}\operatorname{Im}\langle\Psi_{L}|\Phi\rangle
\label{eq:lit_xilin}
\end{equation}
which can be estimated in a numerically stable manner by again sampling from \(\pi_{\Phi}(X)\). 

In addition to providing accurate estimates of the LIT, we provide upper bounds on its uncertainty, whose detailed derivation is given in the Supplemental Material:
\begin{align}
\Delta\mathcal{L}(\omega_{0},\Gamma) &\leq 
\mathcal{D} \; \frac{\mathcal{N}^{-1}||\Phi\rangle|}{\Gamma} \sqrt{\frac{1 - \mathcal{F}  }{ \mathcal{F} }},
\label{eq:error}
\end{align}
where
\begin{equation}
    \mathcal{D} = \min\Big(
\left|(1 - P_{\Phi})|\Psi_L\rangle\right|,\Big|(1 - P_{\Phi})  \frac{H}{\sqrt{\sigma^2 + \Gamma^2}}|\Psi_L\rangle\Big|
\Big),
\end{equation}
and $P_\Phi=|\Phi\rangle\langle \Phi|/\langle \Phi|\Phi\rangle$. This relation provides an important convergence criterion for the quality of the estimated LIT wave function. Noting that in the limit $\Gamma\to 0$, $\omega_0\approx\omega$, and
$\Lag \to (\pi/\Gamma) {\cal R}$, we may conclude that
$\Delta \Lag/\Lag \propto \sqrt{(1-\calF)/\Gamma} $.
Implying that smaller values of $\Gamma$ not only lead to a more complicated wave function, but also demand higher accuracy in solving the LIT equation.

Once the values of the LIT and its uncertainties are estimated on a grid of $N$ discrete values of $\omega_0$, we need to invert the integral transform to retrieve the response function. To this aim, we employ two inversion methods. The first is a regularized version of the standard inversion procedure introduced in Ref.~\cite{Efros:2007nq}, which relies on a suitable basis expansion of the response function. The second is an improved version of the so-called Bryan's version of the Maximum Entropy method~\cite{Jarrell:1996rrw}. Both methods are discussed in detail in the Supplemental material. The advantage of the latter method lies in its ability to propagate the uncertainties in $\mathcal{L}(\omega_0, \Gamma)$ into the reconstructed $\calR(\omega)$ using Bayes' theorem.  

\begin{figure}[!t]
    \includegraphics[width=\linewidth]{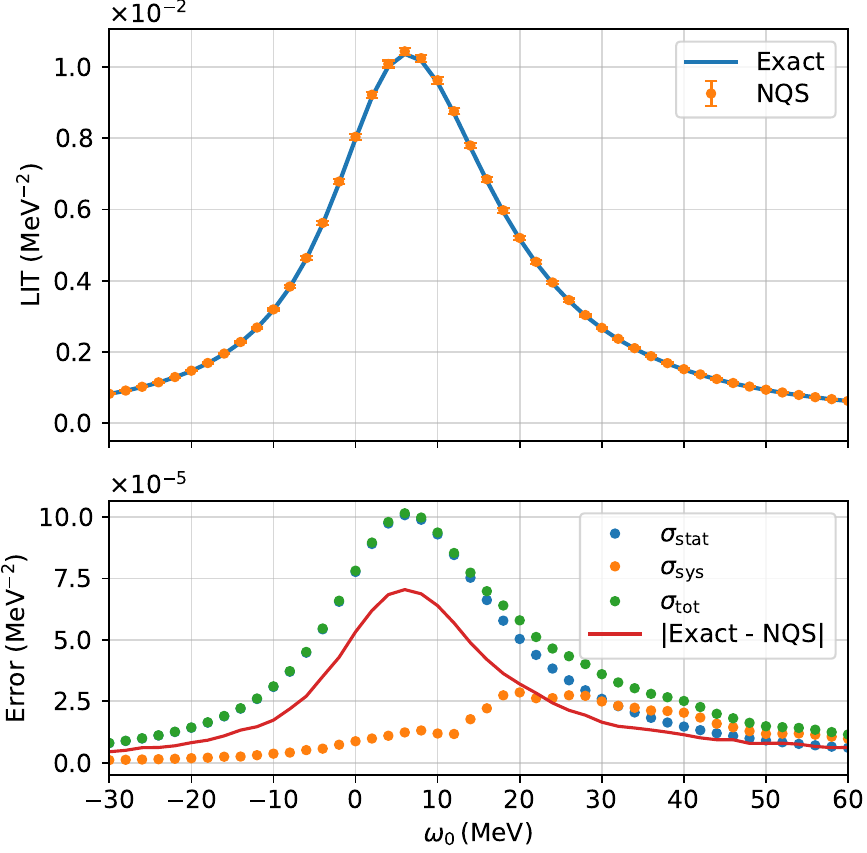}
    \caption{(Upper panel) LIT of the $^2$H dipole response function computed at $\Gamma=10$ MeV. The NQS calculations (orange circles) are compared to the numerically exact results (solid blue line). (Lower panel) The statistical (blue circles), systematic (orange circles), and total (green circles) errors of the NQS calculations are shown alongside the difference between the numerically exact and NQS results (solid red line).} 
    \label{fig:H2_compare}
\end{figure}

\textit{Results.}--- 
Although the method discussed above is entirely general and applicable to a broad range of self-bound systems modeled by NQS in continuum Hilbert spaces, we demonstrate its effectiveness in computing nuclear photoabsorption cross-sections. At intermediate energies, the transition current is well approximated by the translation-invariant dipole operator $\hat O = \sum_i^A q_i\left( z_i- Z_{cm}\right)$, where \( q_i \) is the charge of nucleon $i$, $z_i$ is the $z$ component of its position vector, and $Z_{cm}$ denotes the $z$ component of the center-of-mass vector of the nucleus. 

As a first application, we consider the deuteron, as this simple system allows us to benchmark the NQS-LIT method against a virtually exact numerical solution of the ordinary differential equation that determines the ground-state and LIT wave functions. Since model ``o'' does not include tensor or spin-orbit terms, \( \Psi_0 \) and \( \Psi_L \) contain only the \( s \)-wave and \( p \)-wave contributions, respectively, the differential equations to be solved depend only on the distance between the proton and the neutron. As shown in the upper panel of Fig.~\ref{fig:H2_compare}, the NQS results agree with the exact solution to an accuracy better than 1\% --- both are obtained with $\Gamma=10$ MeV.  This accuracy is compared in the lower panel of Fig.~\ref{fig:H2_compare} with the error bound we determined, and is found to be much better for all values of $\omega$.

\begin{figure}[!b]
    \includegraphics[width=1\linewidth]{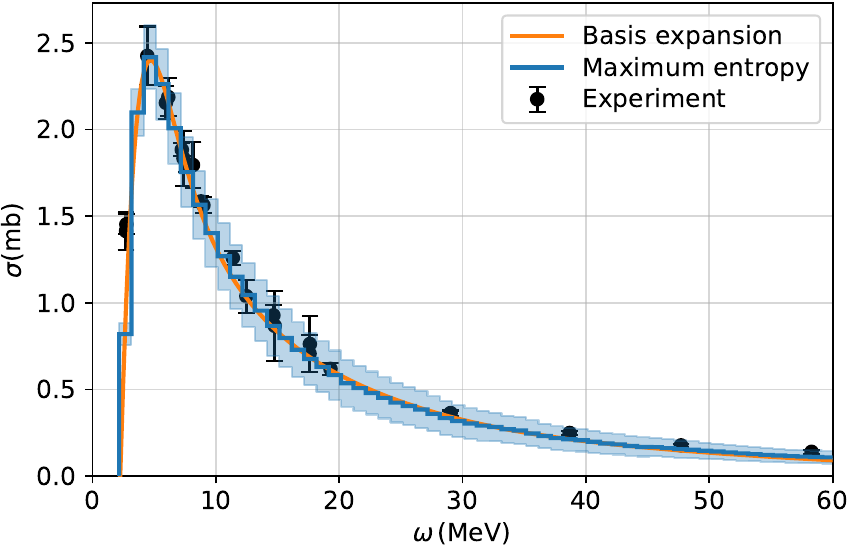}
    \caption{Deuteron photo-disintegration cross section as a function of photon energy, obtained from NQS calculations of the LIT. The basis function results (solid orange lines) and the Maximum Entropy reconstruction (blue histogram with error bars) show good agreement with experimental data from Ref.~\cite{Arenhovel:1990yg}.}
    \label{fig:H2_sigma}
\end{figure}

Inverting the LIT using the basis expansion procedure and maximum-entropy algorithms discussed earlier, we can recover the energy dependence of the response functions and the associated uncertainties. The total photoabsorption cross section is then given by \( \sigma_\gamma(\omega) = 4 \pi^2 \alpha \omega \calR(\omega) \), where \( \alpha \) is the fine-structure constant. In Fig.~\ref{fig:H2_sigma}, we compare the NQS predictions with experimental data from Ref.~\cite{Arenhovel:1990yg}. The basis expansion inversion method shows excellent agreement with the experimental data, regardless of the number \( N_{\text{max}} \) of basis functions used to expand the response—we verified that increasing the basis size up to \( N = 16 \) did not affect the results.
The inversion based on maximum-entropy also yields cross sections that agree well with the data, within uncertainties. This agreement is particularly remarkable given the simplicity of the input Hamiltonian. In this regard, we note that two-body currents are implicitly accounted for by means of the Siegert theorem~\cite{Siegert:1937yt}.

\begin{figure}[!t]
    \includegraphics[width=1\linewidth]{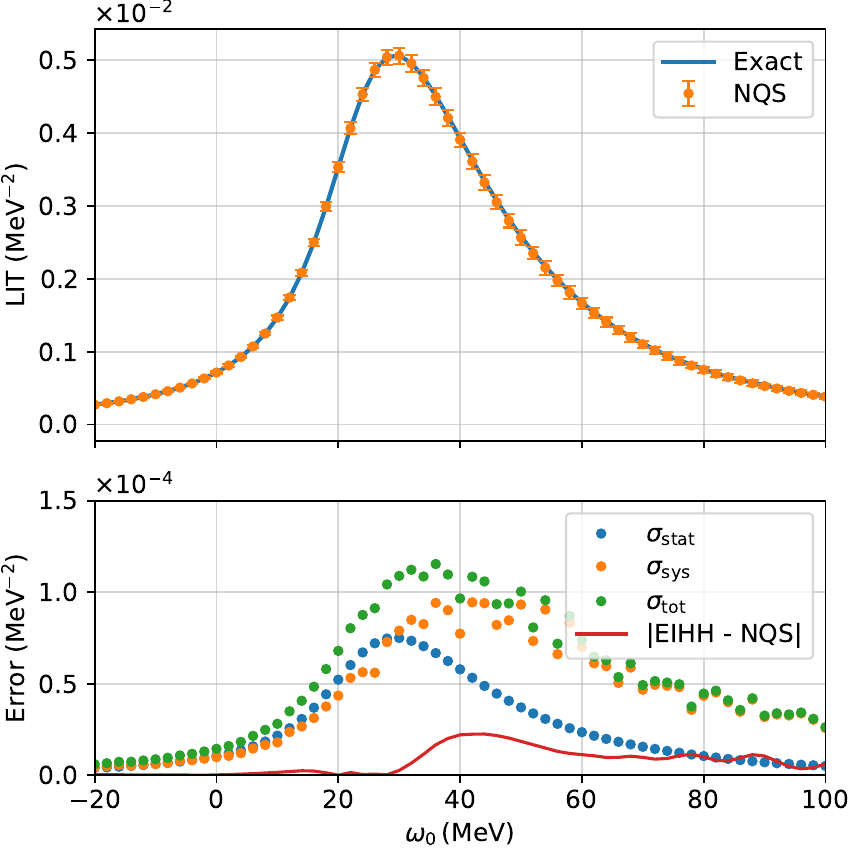}
    \caption{(Upper panel) LIT of the $^4$He dipole response function computed at $\Gamma=10$ MeV. The NQS calculations (orange circles) are compared to the highly-accurate EIHH results (solid blue line). (Lower panel) The statistical (blue circles), systematic (orange circles), and total (green circles) errors of the NQS calculations are shown alongside the difference between the EIHH and NQS results (solid red line).} 
    \label{fig:He4_compare}
\end{figure}

In Fig.~\ref{fig:He4_compare}, we compare the LITs of $^4$He computed using the NQS approach with those obtained via the effective interaction hyperspherical harmonics (EIHH) method~\cite{Barnea:1999be}, which provides extremely accurate solutions for $A\leq 4$ nuclei. Similar to the deuteron case, excellent agreement is also achieved for $^4$He. A detailed analysis of the errors, shown in the lower panel of Fig.~\ref{fig:He4_compare}, reveals that the maximum discrepancy between NQS and EIHH remains at the percent level, except in the high-energy region of the spectrum, where the LIT is already small. Our theoretical error bound is significantly looser, exceeding 20\% for certain $\omega \gtrsim 75$ MeV, indicating the need for refined error estimates.

\begin{figure}[!b]
    \includegraphics[width=1\linewidth]{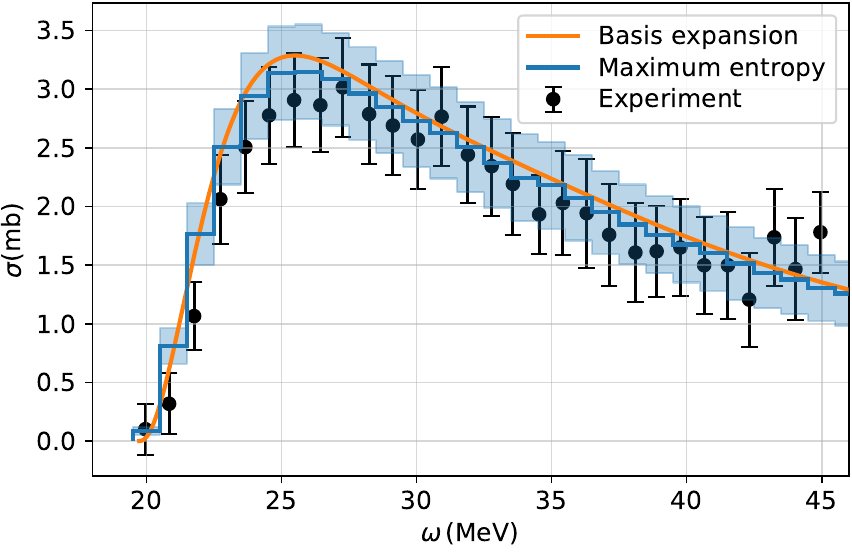}
    \caption{Photo-disintegration cross section of $^4$He as a function of photon energy, obtained from NQS calculations of the LIT. The basis function results (solid orange lines) and the Maximum Entropy reconstruction (blue histogram with error bars) show good agreement with experimental data from Ref.~\cite{Bacca:2014tla}.}
    \label{fig:He4_sigma}
\end{figure}

Finally, in Fig.\ref{fig:He4_sigma}, we present the photodisintegration cross section of $^4$He, obtained by inverting the LIT computed within the NQS framework, and compare it with experimental data from Ref.\cite{Bacca:2014tla}. Similar to the deuteron case, both the basis-expansion inversion method and the maximum entropy approach yield results that agree remarkably well with experiment, especially considering the simplicity of the input Hamiltonian. In fact, the agreement appears even better than that achieved in Refs.~\cite{Bacca:2014rta,Miorelli:2016qbk}, which used higher-resolution chiral Hamiltonians. Notably, the position of the peak is accurately reproduced without the need to shift the theoretical curves to match the experimental threshold. We note that these cross sections were obtained with $\Gamma = 10$ MeV; however, we explicitly verified that using $\Gamma = 5$ MeV produces fully compatible results, with the basis-expansion inversion curves overlapping.

\textit{Conclusions.}---
In this work, we introduced a computational protocol that combines variational Monte Carlo methods based on NQS with the LIT technique to address the challenging problem of computing dynamical properties of self-bound systems in continuous spaces. While this framework is applicable to quantum many-body systems spanning a wide range of energy scales—including atoms and molecules—our initial focus is on applications in nuclear physics. Our approach extends the reach of high-resolution continuum QMC methods to low-energy response functions, including those relevant for photodisintegration cross sections, which until now have only been computed using basis-expansion methods~\cite{Miorelli:2016qbk,Bonaiti:2021kkp,Sobczyk:2021dwm,Bonaiti:2024fft}.

In addition to providing accurate estimates of the response functions, our protocol enables a reliable assessment of both statistical and systematic uncertainties. The statistical uncertainties can be readily estimated through Monte Carlo sampling. To address the systematic component, we derived a general error-bound lemma applicable to all LIT-based methods. We validated this bound numerically through comparisons with highly accurate few-body calculations. This analysis revealed that the upper bound on the error can occasionally be overly conservative, underscoring the need for further refinement of the theoretical estimates.

Both the statistical and systematic uncertainties of the LIT are propagated to the response function by leveraging an advanced version of the Maximum Entropy method. Specifically, the full posterior distribution of the reconstructed spectrum is sampled using a Metropolis algorithm, avoiding the Gaussian approximations employed in earlier applications~\cite{Jarrell:1996rrw}. Our computed photodisintegration cross sections for $^2$H and $^4$He are in excellent agreement with available experimental data, despite the simplicity of the input interaction.

Extending the present approach to nuclei with up to $A=20$ nucleons does not present conceptual difficulties. However, obtaining accurate dipole cross sections—and even the corresponding sum rules—requires precise charge radii. The model ``o'' employed in this work fails to reproduce these observables for nuclei with $A>4$~\cite{Gnech:2024qru}, indicating the need to include $p$-wave components in the nucleon–nucleon interaction. Work in this direction is currently underway.

\textit{Acknowledgments}---A.~L. is deeply grateful to L. Gravina and F. Vicentini for illuminating discussions. The present research is supported by the U.S. Department of Energy, Office of Science, Office of Nuclear Physics, under contracts DE-AC02-06CH11357 (A.~L.), by the DOE Early Career Research Program (A.~L.), by the Fermi Research Alliance, LLC under Contract No. DE-AC02-07CH11359 with the U.S. Department of Energy, Office of Science, Office of High Energy Physics (N.~R.), by the SciDAC-5 NeuCol program (A.~L., N.~R.), and  by the U.S. Department of Energy, Office of Science, Office of Nuclear Physics, under the FRIB Theory Alliance Award No. DE-SC0013617 (X.~Z.) and under the  STREAMLINE Collaboration Award No. DE-SC0024586 (A.~L., N.~R., and X.~Z.). 
This research used resources of the Laboratory Computing Resource Center of Argonne National Laboratory.
E.P, and N.B. would like to thank the Israel Science Foundation
for its support under the grant ISF 2441/24.
\bibliography{biblio}  % Use bibs.bib as the bibliography source

\begin{thebibliography}{53}%
\makeatletter
\providecommand \@ifxundefined [1]{%
 \@ifx{#1\undefined}
}%
\providecommand \@ifnum [1]{%
 \ifnum #1\expandafter \@firstoftwo
 \else \expandafter \@secondoftwo
 \fi
}%
\providecommand \@ifx [1]{%
 \ifx #1\expandafter \@firstoftwo
 \else \expandafter \@secondoftwo
 \fi
}%
\providecommand \natexlab [1]{#1}%
\providecommand \enquote  [1]{``#1''}%
\providecommand \bibnamefont  [1]{#1}%
\providecommand \bibfnamefont [1]{#1}%
\providecommand \citenamefont [1]{#1}%
\providecommand \href@noop [0]{\@secondoftwo}%
\providecommand \href [0]{\begingroup \@sanitize@url \@href}%
\providecommand \@href[1]{\@@startlink{#1}\@@href}%
\providecommand \@@href[1]{\endgroup#1\@@endlink}%
\providecommand \@sanitize@url [0]{\catcode `\\12\catcode `\$12\catcode
  `\&12\catcode `\#12\catcode `\^12\catcode `\_12\catcode `\%12\relax}%
\providecommand \@@startlink[1]{}%
\providecommand \@@endlink[0]{}%
\providecommand \url  [0]{\begingroup\@sanitize@url \@url }%
\providecommand \@url [1]{\endgroup\@href {#1}{\urlprefix }}%
\providecommand \urlprefix  [0]{URL }%
\providecommand \Eprint [0]{\href }%
\providecommand \doibase [0]{http://dx.doi.org/}%
\providecommand \selectlanguage [0]{\@gobble}%
\providecommand \bibinfo  [0]{\@secondoftwo}%
\providecommand \bibfield  [0]{\@secondoftwo}%
\providecommand \translation [1]{[#1]}%
\providecommand \BibitemOpen [0]{}%
\providecommand \bibitemStop [0]{}%
\providecommand \bibitemNoStop [0]{.\EOS\space}%
\providecommand \EOS [0]{\spacefactor3000\relax}%
\providecommand \BibitemShut  [1]{\csname bibitem#1\endcsname}%
\let\auto@bib@innerbib\@empty
%</preamble>
\bibitem [{\citenamefont {Hergert}(2020)}]{Hergert:2020bxy}%
  \BibitemOpen
  \bibfield  {author} {\bibinfo {author} {\bibfnamefont {H.}~\bibnamefont
  {Hergert}},\ }\bibfield  {title} {\enquote {\bibinfo {title} {{A Guided Tour
  of $ab$ $initio$ Nuclear Many-Body Theory}},}\ }\href {\doibase
  10.3389/fphy.2020.00379} {\bibfield  {journal} {\bibinfo  {journal} {Front.
  in Phys.}\ }\textbf {\bibinfo {volume} {8}},\ \bibinfo {pages} {379}
  (\bibinfo {year} {2020})},\ \Eprint {http://arxiv.org/abs/2008.05061}
  {arXiv:2008.05061 [nucl-th]} \BibitemShut {NoStop}%
\bibitem [{\citenamefont {Hagen}\ \emph {et~al.}(2016)\citenamefont {Hagen},
  \citenamefont {Jansen},\ and\ \citenamefont {Papenbrock}}]{Hagen:2016uwj}%
  \BibitemOpen
  \bibfield  {author} {\bibinfo {author} {\bibfnamefont {G.}~\bibnamefont
  {Hagen}}, \bibinfo {author} {\bibfnamefont {G.~R.}\ \bibnamefont {Jansen}}, \
  and\ \bibinfo {author} {\bibfnamefont {T.}~\bibnamefont {Papenbrock}},\
  }\bibfield  {title} {\enquote {\bibinfo {title} {{Structure of $^{78}$Ni from
  first principles computations}},}\ }\href {\doibase
  10.1103/PhysRevLett.117.172501} {\bibfield  {journal} {\bibinfo  {journal}
  {Phys. Rev. Lett.}\ }\textbf {\bibinfo {volume} {117}},\ \bibinfo {pages}
  {172501} (\bibinfo {year} {2016})},\ \Eprint
  {http://arxiv.org/abs/1605.01477} {arXiv:1605.01477 [nucl-th]} \BibitemShut
  {NoStop}%
\bibitem [{\citenamefont {Morris}\ \emph {et~al.}(2018)\citenamefont {Morris},
  \citenamefont {Simonis}, \citenamefont {Stroberg}, \citenamefont {Stumpf},
  \citenamefont {Hagen}, \citenamefont {Holt}, \citenamefont {Jansen},
  \citenamefont {Papenbrock}, \citenamefont {Roth},\ and\ \citenamefont
  {Schwenk}}]{Morris:2017vxi}%
  \BibitemOpen
  \bibfield  {author} {\bibinfo {author} {\bibfnamefont {T.~D.}\ \bibnamefont
  {Morris}}, \bibinfo {author} {\bibfnamefont {J.}~\bibnamefont {Simonis}},
  \bibinfo {author} {\bibfnamefont {S.~R.}\ \bibnamefont {Stroberg}}, \bibinfo
  {author} {\bibfnamefont {C.}~\bibnamefont {Stumpf}}, \bibinfo {author}
  {\bibfnamefont {G.}~\bibnamefont {Hagen}}, \bibinfo {author} {\bibfnamefont
  {J.~D.}\ \bibnamefont {Holt}}, \bibinfo {author} {\bibfnamefont {G.~R.}\
  \bibnamefont {Jansen}}, \bibinfo {author} {\bibfnamefont {T.}~\bibnamefont
  {Papenbrock}}, \bibinfo {author} {\bibfnamefont {R.}~\bibnamefont {Roth}}, \
  and\ \bibinfo {author} {\bibfnamefont {A.}~\bibnamefont {Schwenk}},\
  }\bibfield  {title} {\enquote {\bibinfo {title} {{Structure of the lightest
  tin isotopes}},}\ }\href {\doibase 10.1103/PhysRevLett.120.152503} {\bibfield
   {journal} {\bibinfo  {journal} {Phys. Rev. Lett.}\ }\textbf {\bibinfo
  {volume} {120}},\ \bibinfo {pages} {152503} (\bibinfo {year} {2018})},\
  \Eprint {http://arxiv.org/abs/1709.02786} {arXiv:1709.02786 [nucl-th]}
  \BibitemShut {NoStop}%
\bibitem [{\citenamefont {Gysbers}\ \emph {et~al.}(2019)\citenamefont {Gysbers}
  \emph {et~al.}}]{Gysbers:2019uyb}%
  \BibitemOpen
  \bibfield  {author} {\bibinfo {author} {\bibfnamefont {P.}~\bibnamefont
  {Gysbers}} \emph {et~al.},\ }\bibfield  {title} {\enquote {\bibinfo {title}
  {{Discrepancy between experimental and theoretical $\beta$-decay rates
  resolved from first principles}},}\ }\href {\doibase
  10.1038/s41567-019-0450-7} {\bibfield  {journal} {\bibinfo  {journal} {Nature
  Phys.}\ }\textbf {\bibinfo {volume} {15}},\ \bibinfo {pages} {428--431}
  (\bibinfo {year} {2019})},\ \Eprint {http://arxiv.org/abs/1903.00047}
  {arXiv:1903.00047 [nucl-th]} \BibitemShut {NoStop}%
\bibitem [{\citenamefont {Hu}\ \emph {et~al.}(2022)\citenamefont {Hu} \emph
  {et~al.}}]{Hu:2021trw}%
  \BibitemOpen
  \bibfield  {author} {\bibinfo {author} {\bibfnamefont {Baishan}\ \bibnamefont
  {Hu}} \emph {et~al.},\ }\bibfield  {title} {\enquote {\bibinfo {title} {{Ab
  initio predictions link the neutron skin of $^{208}$Pb to nuclear forces}},}\
  }\href {\doibase 10.1038/s41567-023-02324-9} {\bibfield  {journal} {\bibinfo
  {journal} {Nature Phys.}\ }\textbf {\bibinfo {volume} {18}},\ \bibinfo
  {pages} {1196--1200} (\bibinfo {year} {2022})},\ \Eprint
  {http://arxiv.org/abs/2112.01125} {arXiv:2112.01125 [nucl-th]} \BibitemShut
  {NoStop}%
\bibitem [{\citenamefont {Bacca}\ and\ \citenamefont
  {Pastore}(2014)}]{Bacca:2014tla}%
  \BibitemOpen
  \bibfield  {author} {\bibinfo {author} {\bibfnamefont {Sonia}\ \bibnamefont
  {Bacca}}\ and\ \bibinfo {author} {\bibfnamefont {Saori}\ \bibnamefont
  {Pastore}},\ }\bibfield  {title} {\enquote {\bibinfo {title}
  {{Electromagnetic reactions on light nuclei}},}\ }\href {\doibase
  10.1088/0954-3899/41/12/123002} {\bibfield  {journal} {\bibinfo  {journal}
  {J. Phys. G}\ }\textbf {\bibinfo {volume} {41}},\ \bibinfo {pages} {123002}
  (\bibinfo {year} {2014})},\ \Eprint {http://arxiv.org/abs/1407.3490}
  {arXiv:1407.3490 [nucl-th]} \BibitemShut {NoStop}%
\bibitem [{\citenamefont {Lovato}\ \emph {et~al.}(2013)\citenamefont {Lovato},
  \citenamefont {Gandolfi}, \citenamefont {Butler}, \citenamefont {Carlson},
  \citenamefont {Lusk}, \citenamefont {Pieper},\ and\ \citenamefont
  {Schiavilla}}]{Lovato:2013cua}%
  \BibitemOpen
  \bibfield  {author} {\bibinfo {author} {\bibfnamefont {A.}~\bibnamefont
  {Lovato}}, \bibinfo {author} {\bibfnamefont {S.}~\bibnamefont {Gandolfi}},
  \bibinfo {author} {\bibfnamefont {Ralph}\ \bibnamefont {Butler}}, \bibinfo
  {author} {\bibfnamefont {J.}~\bibnamefont {Carlson}}, \bibinfo {author}
  {\bibfnamefont {Ewing}\ \bibnamefont {Lusk}}, \bibinfo {author}
  {\bibfnamefont {Steven~C.}\ \bibnamefont {Pieper}}, \ and\ \bibinfo {author}
  {\bibfnamefont {R.}~\bibnamefont {Schiavilla}},\ }\bibfield  {title}
  {\enquote {\bibinfo {title} {{Charge Form Factor and Sum Rules of
  Electromagnetic Response Functions in $^{12}C$}},}\ }\href {\doibase
  10.1103/PhysRevLett.111.092501} {\bibfield  {journal} {\bibinfo  {journal}
  {Phys. Rev. Lett.}\ }\textbf {\bibinfo {volume} {111}},\ \bibinfo {pages}
  {092501} (\bibinfo {year} {2013})},\ \Eprint {http://arxiv.org/abs/1305.6959}
  {arXiv:1305.6959 [nucl-th]} \BibitemShut {NoStop}%
\bibitem [{\citenamefont {Clo\"et}\ \emph {et~al.}(2016)\citenamefont
  {Clo\"et}, \citenamefont {Bentz},\ and\ \citenamefont
  {Thomas}}]{Cloet:2015tha}%
  \BibitemOpen
  \bibfield  {author} {\bibinfo {author} {\bibfnamefont {Ian~C.}\ \bibnamefont
  {Clo\"et}}, \bibinfo {author} {\bibfnamefont {Wolfgang}\ \bibnamefont
  {Bentz}}, \ and\ \bibinfo {author} {\bibfnamefont {Anthony~W.}\ \bibnamefont
  {Thomas}},\ }\bibfield  {title} {\enquote {\bibinfo {title} {{Relativistic
  and Nuclear Medium Effects on the Coulomb Sum Rule}},}\ }\href {\doibase
  10.1103/PhysRevLett.116.032701} {\bibfield  {journal} {\bibinfo  {journal}
  {Phys. Rev. Lett.}\ }\textbf {\bibinfo {volume} {116}},\ \bibinfo {pages}
  {032701} (\bibinfo {year} {2016})},\ \Eprint
  {http://arxiv.org/abs/1506.05875} {arXiv:1506.05875 [nucl-th]} \BibitemShut
  {NoStop}%
\bibitem [{\citenamefont {Lovato}\ \emph {et~al.}(2016)\citenamefont {Lovato},
  \citenamefont {Gandolfi}, \citenamefont {Carlson}, \citenamefont {Pieper},\
  and\ \citenamefont {Schiavilla}}]{Lovato:2016gkq}%
  \BibitemOpen
  \bibfield  {author} {\bibinfo {author} {\bibfnamefont {A.}~\bibnamefont
  {Lovato}}, \bibinfo {author} {\bibfnamefont {S.}~\bibnamefont {Gandolfi}},
  \bibinfo {author} {\bibfnamefont {J.}~\bibnamefont {Carlson}}, \bibinfo
  {author} {\bibfnamefont {Steven~C.}\ \bibnamefont {Pieper}}, \ and\ \bibinfo
  {author} {\bibfnamefont {R.}~\bibnamefont {Schiavilla}},\ }\bibfield  {title}
  {\enquote {\bibinfo {title} {{Electromagnetic response of $^{12}$C: A
  first-principles calculation}},}\ }\href {\doibase
  10.1103/PhysRevLett.117.082501} {\bibfield  {journal} {\bibinfo  {journal}
  {Phys. Rev. Lett.}\ }\textbf {\bibinfo {volume} {117}},\ \bibinfo {pages}
  {082501} (\bibinfo {year} {2016})},\ \Eprint
  {http://arxiv.org/abs/1605.00248} {arXiv:1605.00248 [nucl-th]} \BibitemShut
  {NoStop}%
\bibitem [{\citenamefont {Acharya}\ \emph {et~al.}(2024)\citenamefont
  {Acharya}, \citenamefont {Sobczyk}, \citenamefont {Bacca}, \citenamefont
  {Hagen},\ and\ \citenamefont {Jiang}}]{Acharya:2024xah}%
  \BibitemOpen
  \bibfield  {author} {\bibinfo {author} {\bibfnamefont {Bijaya}\ \bibnamefont
  {Acharya}}, \bibinfo {author} {\bibfnamefont {Joanna~E.}\ \bibnamefont
  {Sobczyk}}, \bibinfo {author} {\bibfnamefont {Sonia}\ \bibnamefont {Bacca}},
  \bibinfo {author} {\bibfnamefont {Gaute}\ \bibnamefont {Hagen}}, \ and\
  \bibinfo {author} {\bibfnamefont {Weiguang}\ \bibnamefont {Jiang}},\
  }\bibfield  {title} {\enquote {\bibinfo {title} {{16O electroweak response
  functions from first principles}},}\ }\href@noop {} {\  (\bibinfo {year}
  {2024})},\ \Eprint {http://arxiv.org/abs/2410.05962} {arXiv:2410.05962
  [nucl-th]} \BibitemShut {NoStop}%
\bibitem [{\citenamefont {Lovato}\ \emph {et~al.}(2020)\citenamefont {Lovato},
  \citenamefont {Carlson}, \citenamefont {Gandolfi}, \citenamefont {Rocco},\
  and\ \citenamefont {Schiavilla}}]{Lovato:2020kba}%
  \BibitemOpen
  \bibfield  {author} {\bibinfo {author} {\bibfnamefont {A.}~\bibnamefont
  {Lovato}}, \bibinfo {author} {\bibfnamefont {J.}~\bibnamefont {Carlson}},
  \bibinfo {author} {\bibfnamefont {S.}~\bibnamefont {Gandolfi}}, \bibinfo
  {author} {\bibfnamefont {N.}~\bibnamefont {Rocco}}, \ and\ \bibinfo {author}
  {\bibfnamefont {R.}~\bibnamefont {Schiavilla}},\ }\bibfield  {title}
  {\enquote {\bibinfo {title} {{Ab initio study of
  $\boldsymbol{(\nu_\ell,\ell^-)}$ and
  $\boldsymbol{(\overline{\nu}_\ell,\ell^+)}$ inclusive scattering in $^{12}$C:
  confronting the MiniBooNE and T2K CCQE data}},}\ }\href {\doibase
  10.1103/PhysRevX.10.031068} {\bibfield  {journal} {\bibinfo  {journal} {Phys.
  Rev. X}\ }\textbf {\bibinfo {volume} {10}},\ \bibinfo {pages} {031068}
  (\bibinfo {year} {2020})},\ \Eprint {http://arxiv.org/abs/2003.07710}
  {arXiv:2003.07710 [nucl-th]} \BibitemShut {NoStop}%
\bibitem [{\citenamefont {Yakovlev}\ and\ \citenamefont
  {Pethick}(2004)}]{Yakovlev:2004iq}%
  \BibitemOpen
  \bibfield  {author} {\bibinfo {author} {\bibfnamefont {Dima~G.}\ \bibnamefont
  {Yakovlev}}\ and\ \bibinfo {author} {\bibfnamefont {C.~J.}\ \bibnamefont
  {Pethick}},\ }\bibfield  {title} {\enquote {\bibinfo {title} {{Neutron star
  cooling}},}\ }\href {\doibase 10.1146/annurev.astro.42.053102.134013}
  {\bibfield  {journal} {\bibinfo  {journal} {Ann. Rev. Astron. Astrophys.}\
  }\textbf {\bibinfo {volume} {42}},\ \bibinfo {pages} {169--210} (\bibinfo
  {year} {2004})},\ \Eprint {http://arxiv.org/abs/astro-ph/0402143}
  {arXiv:astro-ph/0402143} \BibitemShut {NoStop}%
\bibitem [{\citenamefont {Shen}\ \emph {et~al.}(2013)\citenamefont {Shen},
  \citenamefont {Gandolfi}, \citenamefont {Reddy},\ and\ \citenamefont
  {Carlson}}]{Shen:2012sa}%
  \BibitemOpen
  \bibfield  {author} {\bibinfo {author} {\bibfnamefont {G.}~\bibnamefont
  {Shen}}, \bibinfo {author} {\bibfnamefont {S.}~\bibnamefont {Gandolfi}},
  \bibinfo {author} {\bibfnamefont {S.}~\bibnamefont {Reddy}}, \ and\ \bibinfo
  {author} {\bibfnamefont {J.}~\bibnamefont {Carlson}},\ }\bibfield  {title}
  {\enquote {\bibinfo {title} {{Spin response and neutrino emissivity of dense
  neutron matter}},}\ }\href {\doibase 10.1103/PhysRevC.87.025802} {\bibfield
  {journal} {\bibinfo  {journal} {Phys. Rev. C}\ }\textbf {\bibinfo {volume}
  {87}},\ \bibinfo {pages} {025802} (\bibinfo {year} {2013})},\ \Eprint
  {http://arxiv.org/abs/1205.6499} {arXiv:1205.6499 [nucl-th]} \BibitemShut
  {NoStop}%
\bibitem [{\citenamefont {Sobczyk}\ \emph
  {et~al.}(2024{\natexlab{a}})\citenamefont {Sobczyk}, \citenamefont {Jiang},\
  and\ \citenamefont {Roggero}}]{Sobczyk:2024hdl}%
  \BibitemOpen
  \bibfield  {author} {\bibinfo {author} {\bibfnamefont {J.~E.}\ \bibnamefont
  {Sobczyk}}, \bibinfo {author} {\bibfnamefont {W.}~\bibnamefont {Jiang}}, \
  and\ \bibinfo {author} {\bibfnamefont {A.}~\bibnamefont {Roggero}},\
  }\bibfield  {title} {\enquote {\bibinfo {title} {{Spin response of neutron
  matter in ab initio approach}},}\ }\href@noop {} {\bibfield  {journal}
  {\bibinfo  {journal} {Placeholder Journal}\ } (\bibinfo {year}
  {2024}{\natexlab{a}})},\ \Eprint {http://arxiv.org/abs/2407.20986}
  {arXiv:2407.20986 [nucl-th]} \BibitemShut {NoStop}%
\bibitem [{\citenamefont {Carlson}\ \emph {et~al.}(2015)\citenamefont
  {Carlson}, \citenamefont {Gandolfi}, \citenamefont {Pederiva}, \citenamefont
  {Pieper}, \citenamefont {Schiavilla}, \citenamefont {Schmidt},\ and\
  \citenamefont {Wiringa}}]{Carlson:2014vla}%
  \BibitemOpen
  \bibfield  {author} {\bibinfo {author} {\bibfnamefont {J.}~\bibnamefont
  {Carlson}}, \bibinfo {author} {\bibfnamefont {S.}~\bibnamefont {Gandolfi}},
  \bibinfo {author} {\bibfnamefont {F.}~\bibnamefont {Pederiva}}, \bibinfo
  {author} {\bibfnamefont {Steven~C.}\ \bibnamefont {Pieper}}, \bibinfo
  {author} {\bibfnamefont {R.}~\bibnamefont {Schiavilla}}, \bibinfo {author}
  {\bibfnamefont {K.~E.}\ \bibnamefont {Schmidt}}, \ and\ \bibinfo {author}
  {\bibfnamefont {R.~B.}\ \bibnamefont {Wiringa}},\ }\bibfield  {title}
  {\enquote {\bibinfo {title} {{Quantum Monte Carlo methods for nuclear
  physics}},}\ }\href {\doibase 10.1103/RevModPhys.87.1067} {\bibfield
  {journal} {\bibinfo  {journal} {Rev. Mod. Phys.}\ }\textbf {\bibinfo {volume}
  {87}},\ \bibinfo {pages} {1067} (\bibinfo {year} {2015})},\ \Eprint
  {http://arxiv.org/abs/1412.3081} {arXiv:1412.3081 [nucl-th]} \BibitemShut
  {NoStop}%
\bibitem [{\citenamefont {Nikolakopoulos}\ \emph {et~al.}(2024)\citenamefont
  {Nikolakopoulos}, \citenamefont {Lovato},\ and\ \citenamefont
  {Rocco}}]{Nikolakopoulos:2023zse}%
  \BibitemOpen
  \bibfield  {author} {\bibinfo {author} {\bibfnamefont {Alexis}\ \bibnamefont
  {Nikolakopoulos}}, \bibinfo {author} {\bibfnamefont {Alessandro}\
  \bibnamefont {Lovato}}, \ and\ \bibinfo {author} {\bibfnamefont {Noemi}\
  \bibnamefont {Rocco}},\ }\bibfield  {title} {\enquote {\bibinfo {title}
  {{Relativistic effects in Green's function Monte Carlo calculations of
  neutrino-nucleus scattering}},}\ }\href {\doibase
  10.1103/PhysRevC.109.014623} {\bibfield  {journal} {\bibinfo  {journal}
  {Phys. Rev. C}\ }\textbf {\bibinfo {volume} {109}},\ \bibinfo {pages}
  {014623} (\bibinfo {year} {2024})},\ \Eprint
  {http://arxiv.org/abs/2304.11772} {arXiv:2304.11772 [nucl-th]} \BibitemShut
  {NoStop}%
\bibitem [{\citenamefont {Gnech}\ \emph {et~al.}(2025)\citenamefont {Gnech},
  \citenamefont {Lovato},\ and\ \citenamefont {Rocco}}]{Gnech:2024qru}%
  \BibitemOpen
  \bibfield  {author} {\bibinfo {author} {\bibfnamefont {Alex}\ \bibnamefont
  {Gnech}}, \bibinfo {author} {\bibfnamefont {Alessandro}\ \bibnamefont
  {Lovato}}, \ and\ \bibinfo {author} {\bibfnamefont {Noemi}\ \bibnamefont
  {Rocco}},\ }\bibfield  {title} {\enquote {\bibinfo {title} {{Static and
  dynamic properties of atomic nuclei with high-resolution potentials}},}\
  }\href {\doibase 10.1103/PhysRevC.111.024314} {\bibfield  {journal} {\bibinfo
   {journal} {Phys. Rev. C}\ }\textbf {\bibinfo {volume} {111}},\ \bibinfo
  {pages} {024314} (\bibinfo {year} {2025})},\ \Eprint
  {http://arxiv.org/abs/2405.14916} {arXiv:2405.14916 [nucl-th]} \BibitemShut
  {NoStop}%
\bibitem [{\citenamefont {Raghavan}\ \emph {et~al.}(2021)\citenamefont
  {Raghavan}, \citenamefont {Balaprakash}, \citenamefont {Lovato},
  \citenamefont {Rocco},\ and\ \citenamefont {Wild}}]{Raghavan:2020bze}%
  \BibitemOpen
  \bibfield  {author} {\bibinfo {author} {\bibfnamefont {Krishnan}\
  \bibnamefont {Raghavan}}, \bibinfo {author} {\bibfnamefont {Prasanna}\
  \bibnamefont {Balaprakash}}, \bibinfo {author} {\bibfnamefont {Alessandro}\
  \bibnamefont {Lovato}}, \bibinfo {author} {\bibfnamefont {Noemi}\
  \bibnamefont {Rocco}}, \ and\ \bibinfo {author} {\bibfnamefont {Stefan~M.}\
  \bibnamefont {Wild}},\ }\bibfield  {title} {\enquote {\bibinfo {title}
  {{Machine learning-based inversion of nuclear responses}},}\ }\href {\doibase
  10.1103/PhysRevC.103.035502} {\bibfield  {journal} {\bibinfo  {journal}
  {Phys. Rev. C}\ }\textbf {\bibinfo {volume} {103}},\ \bibinfo {pages}
  {035502} (\bibinfo {year} {2021})},\ \Eprint
  {http://arxiv.org/abs/2010.12703} {arXiv:2010.12703 [nucl-th]} \BibitemShut
  {NoStop}%
\bibitem [{\citenamefont {Raghavan}\ and\ \citenamefont
  {Lovato}(2024)}]{Raghavan:2023pav}%
  \BibitemOpen
  \bibfield  {author} {\bibinfo {author} {\bibfnamefont {Krishnan}\
  \bibnamefont {Raghavan}}\ and\ \bibinfo {author} {\bibfnamefont {Alessandro}\
  \bibnamefont {Lovato}},\ }\bibfield  {title} {\enquote {\bibinfo {title}
  {{Uncertainty-quantification-enabled inversion of nuclear responses}},}\
  }\href {\doibase 10.1103/PhysRevC.110.025504} {\bibfield  {journal} {\bibinfo
   {journal} {Phys. Rev. C}\ }\textbf {\bibinfo {volume} {110}},\ \bibinfo
  {pages} {025504} (\bibinfo {year} {2024})},\ \Eprint
  {http://arxiv.org/abs/2310.18756} {arXiv:2310.18756 [nucl-th]} \BibitemShut
  {NoStop}%
\bibitem [{\citenamefont {Hagen}\ \emph {et~al.}(2014)\citenamefont {Hagen},
  \citenamefont {Papenbrock}, \citenamefont {Hjorth-Jensen},\ and\
  \citenamefont {Dean}}]{Hagen:2013nca}%
  \BibitemOpen
  \bibfield  {author} {\bibinfo {author} {\bibfnamefont {G.}~\bibnamefont
  {Hagen}}, \bibinfo {author} {\bibfnamefont {T.}~\bibnamefont {Papenbrock}},
  \bibinfo {author} {\bibfnamefont {M.}~\bibnamefont {Hjorth-Jensen}}, \ and\
  \bibinfo {author} {\bibfnamefont {D.~J.}\ \bibnamefont {Dean}},\ }\bibfield
  {title} {\enquote {\bibinfo {title} {{Coupled-cluster computations of atomic
  nuclei}},}\ }\href {\doibase 10.1088/0034-4885/77/9/096302} {\bibfield
  {journal} {\bibinfo  {journal} {Rept. Prog. Phys.}\ }\textbf {\bibinfo
  {volume} {77}},\ \bibinfo {pages} {096302} (\bibinfo {year} {2014})},\
  \Eprint {http://arxiv.org/abs/1312.7872} {arXiv:1312.7872 [nucl-th]}
  \BibitemShut {NoStop}%
\bibitem [{\citenamefont {Sobczyk}\ \emph {et~al.}(2021)\citenamefont
  {Sobczyk}, \citenamefont {Acharya}, \citenamefont {Bacca},\ and\
  \citenamefont {Hagen}}]{Sobczyk:2021dwm}%
  \BibitemOpen
  \bibfield  {author} {\bibinfo {author} {\bibfnamefont {J.~E.}\ \bibnamefont
  {Sobczyk}}, \bibinfo {author} {\bibfnamefont {B.}~\bibnamefont {Acharya}},
  \bibinfo {author} {\bibfnamefont {S.}~\bibnamefont {Bacca}}, \ and\ \bibinfo
  {author} {\bibfnamefont {G.}~\bibnamefont {Hagen}},\ }\bibfield  {title}
  {\enquote {\bibinfo {title} {{Ab initio computation of the longitudinal
  response function in $^{40}$Ca}},}\ }\href {\doibase
  10.1103/PhysRevLett.127.072501} {\bibfield  {journal} {\bibinfo  {journal}
  {Phys. Rev. Lett.}\ }\textbf {\bibinfo {volume} {127}},\ \bibinfo {pages}
  {072501} (\bibinfo {year} {2021})},\ \Eprint
  {http://arxiv.org/abs/2103.06786} {arXiv:2103.06786 [nucl-th]} \BibitemShut
  {NoStop}%
\bibitem [{\citenamefont {Sobczyk}\ \emph
  {et~al.}(2024{\natexlab{b}})\citenamefont {Sobczyk}, \citenamefont {Acharya},
  \citenamefont {Bacca},\ and\ \citenamefont {Hagen}}]{Sobczyk:2023sxh}%
  \BibitemOpen
  \bibfield  {author} {\bibinfo {author} {\bibfnamefont {J.~E.}\ \bibnamefont
  {Sobczyk}}, \bibinfo {author} {\bibfnamefont {B.}~\bibnamefont {Acharya}},
  \bibinfo {author} {\bibfnamefont {S.}~\bibnamefont {Bacca}}, \ and\ \bibinfo
  {author} {\bibfnamefont {G.}~\bibnamefont {Hagen}},\ }\bibfield  {title}
  {\enquote {\bibinfo {title} {{Ca40 transverse response function from
  coupled-cluster theory}},}\ }\href {\doibase 10.1103/PhysRevC.109.025502}
  {\bibfield  {journal} {\bibinfo  {journal} {Phys. Rev. C}\ }\textbf {\bibinfo
  {volume} {109}},\ \bibinfo {pages} {025502} (\bibinfo {year}
  {2024}{\natexlab{b}})},\ \Eprint {http://arxiv.org/abs/2310.03109}
  {arXiv:2310.03109 [nucl-th]} \BibitemShut {NoStop}%
\bibitem [{\citenamefont {Miorelli}\ \emph {et~al.}(2016)\citenamefont
  {Miorelli}, \citenamefont {Bacca}, \citenamefont {Barnea}, \citenamefont
  {Hagen}, \citenamefont {Jansen}, \citenamefont {Orlandini},\ and\
  \citenamefont {Papenbrock}}]{Miorelli:2016qbk}%
  \BibitemOpen
  \bibfield  {author} {\bibinfo {author} {\bibfnamefont {M.}~\bibnamefont
  {Miorelli}}, \bibinfo {author} {\bibfnamefont {S.}~\bibnamefont {Bacca}},
  \bibinfo {author} {\bibfnamefont {N.}~\bibnamefont {Barnea}}, \bibinfo
  {author} {\bibfnamefont {G.}~\bibnamefont {Hagen}}, \bibinfo {author}
  {\bibfnamefont {G.~R.}\ \bibnamefont {Jansen}}, \bibinfo {author}
  {\bibfnamefont {G.}~\bibnamefont {Orlandini}}, \ and\ \bibinfo {author}
  {\bibfnamefont {T.}~\bibnamefont {Papenbrock}},\ }\bibfield  {title}
  {\enquote {\bibinfo {title} {{Electric dipole polarizability from first
  principles calculations}},}\ }\href {\doibase 10.1103/PhysRevC.94.034317}
  {\bibfield  {journal} {\bibinfo  {journal} {Phys. Rev. C}\ }\textbf {\bibinfo
  {volume} {94}},\ \bibinfo {pages} {034317} (\bibinfo {year} {2016})},\
  \Eprint {http://arxiv.org/abs/1604.05381} {arXiv:1604.05381 [nucl-th]}
  \BibitemShut {NoStop}%
\bibitem [{\citenamefont {Birkhan}\ \emph {et~al.}(2017)\citenamefont {Birkhan}
  \emph {et~al.}}]{Birkhan:2016qkr}%
  \BibitemOpen
  \bibfield  {author} {\bibinfo {author} {\bibfnamefont {J.}~\bibnamefont
  {Birkhan}} \emph {et~al.},\ }\bibfield  {title} {\enquote {\bibinfo {title}
  {{Electric dipole polarizability of $^{48}$Ca and implications for the
  neutron skin}},}\ }\href {\doibase 10.1103/PhysRevLett.118.252501} {\bibfield
   {journal} {\bibinfo  {journal} {Phys. Rev. Lett.}\ }\textbf {\bibinfo
  {volume} {118}},\ \bibinfo {pages} {252501} (\bibinfo {year} {2017})},\
  \Eprint {http://arxiv.org/abs/1611.07072} {arXiv:1611.07072 [nucl-ex]}
  \BibitemShut {NoStop}%
\bibitem [{\citenamefont {Fearick}\ \emph {et~al.}(2023)\citenamefont {Fearick}
  \emph {et~al.}}]{Fearick:2023lyz}%
  \BibitemOpen
  \bibfield  {author} {\bibinfo {author} {\bibfnamefont {R.~W.}\ \bibnamefont
  {Fearick}} \emph {et~al.},\ }\bibfield  {title} {\enquote {\bibinfo {title}
  {{Electric dipole polarizability of Ca40}},}\ }\href {\doibase
  10.1103/PhysRevResearch.5.L022044} {\bibfield  {journal} {\bibinfo  {journal}
  {Phys. Rev. Res.}\ }\textbf {\bibinfo {volume} {5}},\ \bibinfo {pages}
  {L022044} (\bibinfo {year} {2023})},\ \Eprint
  {http://arxiv.org/abs/2302.07490} {arXiv:2302.07490 [nucl-ex]} \BibitemShut
  {NoStop}%
\bibitem [{\citenamefont {Bonaiti}\ \emph {et~al.}(2022)\citenamefont
  {Bonaiti}, \citenamefont {Bacca},\ and\ \citenamefont
  {Hagen}}]{Bonaiti:2021kkp}%
  \BibitemOpen
  \bibfield  {author} {\bibinfo {author} {\bibfnamefont {Francesca}\
  \bibnamefont {Bonaiti}}, \bibinfo {author} {\bibfnamefont {Sonia}\
  \bibnamefont {Bacca}}, \ and\ \bibinfo {author} {\bibfnamefont {Gaute}\
  \bibnamefont {Hagen}},\ }\bibfield  {title} {\enquote {\bibinfo {title} {{Ab
  initio coupled-cluster calculations of ground and dipole excited states in
  He8}},}\ }\href {\doibase 10.1103/PhysRevC.105.034313} {\bibfield  {journal}
  {\bibinfo  {journal} {Phys. Rev. C}\ }\textbf {\bibinfo {volume} {105}},\
  \bibinfo {pages} {034313} (\bibinfo {year} {2022})},\ \Eprint
  {http://arxiv.org/abs/2112.08210} {arXiv:2112.08210 [nucl-th]} \BibitemShut
  {NoStop}%
\bibitem [{\citenamefont {Bonaiti}\ \emph {et~al.}(2024)\citenamefont
  {Bonaiti}, \citenamefont {Bacca}, \citenamefont {Hagen},\ and\ \citenamefont
  {Jansen}}]{Bonaiti:2024fft}%
  \BibitemOpen
  \bibfield  {author} {\bibinfo {author} {\bibfnamefont {Francesca}\
  \bibnamefont {Bonaiti}}, \bibinfo {author} {\bibfnamefont {Sonia}\
  \bibnamefont {Bacca}}, \bibinfo {author} {\bibfnamefont {Gaute}\ \bibnamefont
  {Hagen}}, \ and\ \bibinfo {author} {\bibfnamefont {Gustav~R.}\ \bibnamefont
  {Jansen}},\ }\bibfield  {title} {\enquote {\bibinfo {title} {{Electromagnetic
  observables of open-shell nuclei from coupled-cluster theory}},}\ }\href
  {\doibase 10.1103/PhysRevC.110.044306} {\bibfield  {journal} {\bibinfo
  {journal} {Phys. Rev. C}\ }\textbf {\bibinfo {volume} {110}},\ \bibinfo
  {pages} {044306} (\bibinfo {year} {2024})},\ \Eprint
  {http://arxiv.org/abs/2405.05608} {arXiv:2405.05608 [nucl-th]} \BibitemShut
  {NoStop}%
\bibitem [{\citenamefont {Burrows}\ \emph {et~al.}(2023)\citenamefont
  {Burrows}, \citenamefont {Baker}, \citenamefont {Bacca}, \citenamefont
  {Launey}, \citenamefont {Dytrych},\ and\ \citenamefont
  {Langr}}]{Burrows:2023ugy}%
  \BibitemOpen
  \bibfield  {author} {\bibinfo {author} {\bibfnamefont {M.}~\bibnamefont
  {Burrows}}, \bibinfo {author} {\bibfnamefont {R.~B.}\ \bibnamefont {Baker}},
  \bibinfo {author} {\bibfnamefont {S.}~\bibnamefont {Bacca}}, \bibinfo
  {author} {\bibfnamefont {K.~D.}\ \bibnamefont {Launey}}, \bibinfo {author}
  {\bibfnamefont {T.}~\bibnamefont {Dytrych}}, \ and\ \bibinfo {author}
  {\bibfnamefont {D.}~\bibnamefont {Langr}},\ }\bibfield  {title} {\enquote
  {\bibinfo {title} {{Response functions and giant monopole resonances for
  light to medium-mass nuclei from the \textbackslash{}textit{ab initio}
  symmetry-adapted no-core shell model}},}\ }\href@noop {} {\  (\bibinfo {year}
  {2023})},\ \Eprint {http://arxiv.org/abs/2312.09782} {arXiv:2312.09782
  [nucl-th]} \BibitemShut {NoStop}%
\bibitem [{\citenamefont {Carleo}\ and\ \citenamefont
  {Troyer}(2017)}]{Carleo:2017}%
  \BibitemOpen
  \bibfield  {author} {\bibinfo {author} {\bibfnamefont {Giuseppe}\
  \bibnamefont {Carleo}}\ and\ \bibinfo {author} {\bibfnamefont {Matthias}\
  \bibnamefont {Troyer}},\ }\bibfield  {title} {\enquote {\bibinfo {title}
  {Solving the quantum many-body problem with artificial neural networks},}\
  }\href {\doibase 10.1126/science.aag2302} {\bibfield  {journal} {\bibinfo
  {journal} {Science}\ }\textbf {\bibinfo {volume} {355}},\ \bibinfo {pages}
  {602--606} (\bibinfo {year} {2017})}\BibitemShut {NoStop}%
\bibitem [{\citenamefont {Hermann}\ \emph {et~al.}(2020)\citenamefont
  {Hermann}, \citenamefont {Sch\"atzle},\ and\ \citenamefont
  {No\'e}}]{Hermann:2020xqs}%
  \BibitemOpen
  \bibfield  {author} {\bibinfo {author} {\bibfnamefont {Jan}\ \bibnamefont
  {Hermann}}, \bibinfo {author} {\bibfnamefont {Zeno}\ \bibnamefont
  {Sch\"atzle}}, \ and\ \bibinfo {author} {\bibfnamefont {Frank}\ \bibnamefont
  {No\'e}},\ }\bibfield  {title} {\enquote {\bibinfo {title}
  {{Deep-neural-network solution of the electronic Schr\"odinger equation}},}\
  }\href {\doibase 10.1038/s41557-020-0544-y} {\bibfield  {journal} {\bibinfo
  {journal} {Nature Chem.}\ }\textbf {\bibinfo {volume} {12}},\ \bibinfo
  {pages} {891--897} (\bibinfo {year} {2020})}\BibitemShut {NoStop}%
\bibitem [{\citenamefont {Pfau}\ \emph {et~al.}(2020)\citenamefont {Pfau},
  \citenamefont {Spencer}, \citenamefont {Matthews},\ and\ \citenamefont
  {Foulkes}}]{Pfau:2020}%
  \BibitemOpen
  \bibfield  {author} {\bibinfo {author} {\bibfnamefont {David}\ \bibnamefont
  {Pfau}}, \bibinfo {author} {\bibfnamefont {James~S.}\ \bibnamefont
  {Spencer}}, \bibinfo {author} {\bibfnamefont {Alexander G. D.~G.}\
  \bibnamefont {Matthews}}, \ and\ \bibinfo {author} {\bibfnamefont {W.~M.~C.}\
  \bibnamefont {Foulkes}},\ }\bibfield  {title} {\enquote {\bibinfo {title} {Ab
  initio solution of the many-electron schr\"odinger equation with deep neural
  networks},}\ }\href {\doibase 10.1103/PhysRevResearch.2.033429} {\bibfield
  {journal} {\bibinfo  {journal} {Phys. Rev. Res.}\ }\textbf {\bibinfo {volume}
  {2}},\ \bibinfo {pages} {033429} (\bibinfo {year} {2020})}\BibitemShut
  {NoStop}%
\bibitem [{\citenamefont {Keeble}\ and\ \citenamefont
  {Rios}(2020)}]{Keeble:2019bkv}%
  \BibitemOpen
  \bibfield  {author} {\bibinfo {author} {\bibfnamefont {J.~W.~T.}\
  \bibnamefont {Keeble}}\ and\ \bibinfo {author} {\bibfnamefont
  {A.}~\bibnamefont {Rios}},\ }\bibfield  {title} {\enquote {\bibinfo {title}
  {{Machine learning the deuteron}},}\ }\href {\doibase
  10.1016/j.physletb.2020.135743} {\bibfield  {journal} {\bibinfo  {journal}
  {Phys. Lett. B}\ }\textbf {\bibinfo {volume} {809}},\ \bibinfo {pages}
  {135743} (\bibinfo {year} {2020})},\ \Eprint
  {http://arxiv.org/abs/1911.13092} {arXiv:1911.13092 [nucl-th]} \BibitemShut
  {NoStop}%
\bibitem [{\citenamefont {Adams}\ \emph {et~al.}(2021)\citenamefont {Adams},
  \citenamefont {Carleo}, \citenamefont {Lovato},\ and\ \citenamefont
  {Rocco}}]{Adams:2020aax}%
  \BibitemOpen
  \bibfield  {author} {\bibinfo {author} {\bibfnamefont {Corey}\ \bibnamefont
  {Adams}}, \bibinfo {author} {\bibfnamefont {Giuseppe}\ \bibnamefont
  {Carleo}}, \bibinfo {author} {\bibfnamefont {Alessandro}\ \bibnamefont
  {Lovato}}, \ and\ \bibinfo {author} {\bibfnamefont {Noemi}\ \bibnamefont
  {Rocco}},\ }\bibfield  {title} {\enquote {\bibinfo {title} {{Variational
  Monte Carlo Calculations of A\ensuremath{\leq}4 Nuclei with an Artificial
  Neural-Network Correlator Ansatz}},}\ }\href {\doibase
  10.1103/PhysRevLett.127.022502} {\bibfield  {journal} {\bibinfo  {journal}
  {Phys. Rev. Lett.}\ }\textbf {\bibinfo {volume} {127}},\ \bibinfo {pages}
  {022502} (\bibinfo {year} {2021})},\ \Eprint
  {http://arxiv.org/abs/2007.14282} {arXiv:2007.14282 [nucl-th]} \BibitemShut
  {NoStop}%
\bibitem [{\citenamefont {Lovato}\ \emph {et~al.}(2022)\citenamefont {Lovato},
  \citenamefont {Adams}, \citenamefont {Carleo},\ and\ \citenamefont
  {Rocco}}]{Lovato:2022tjh}%
  \BibitemOpen
  \bibfield  {author} {\bibinfo {author} {\bibfnamefont {Alessandro}\
  \bibnamefont {Lovato}}, \bibinfo {author} {\bibfnamefont {Corey}\
  \bibnamefont {Adams}}, \bibinfo {author} {\bibfnamefont {Giuseppe}\
  \bibnamefont {Carleo}}, \ and\ \bibinfo {author} {\bibfnamefont {Noemi}\
  \bibnamefont {Rocco}},\ }\bibfield  {title} {\enquote {\bibinfo {title}
  {{Hidden-nucleons neural-network quantum states for the nuclear many-body
  problem}},}\ }\href {\doibase 10.1103/PhysRevResearch.4.043178} {\bibfield
  {journal} {\bibinfo  {journal} {Phys. Rev. Res.}\ }\textbf {\bibinfo {volume}
  {4}},\ \bibinfo {pages} {043178} (\bibinfo {year} {2022})},\ \Eprint
  {http://arxiv.org/abs/2206.10021} {arXiv:2206.10021 [nucl-th]} \BibitemShut
  {NoStop}%
\bibitem [{\citenamefont {Yang}\ and\ \citenamefont
  {Zhao}(2022)}]{Yang:2022esu}%
  \BibitemOpen
  \bibfield  {author} {\bibinfo {author} {\bibfnamefont {Y.~L.}\ \bibnamefont
  {Yang}}\ and\ \bibinfo {author} {\bibfnamefont {P.~W.}\ \bibnamefont
  {Zhao}},\ }\bibfield  {title} {\enquote {\bibinfo {title} {{A consistent
  description of the relativistic effects and three-body interactions in atomic
  nuclei}},}\ }\href {\doibase 10.1016/j.physletb.2022.137587} {\bibfield
  {journal} {\bibinfo  {journal} {Phys. Lett. B}\ }\textbf {\bibinfo {volume}
  {835}},\ \bibinfo {pages} {137587} (\bibinfo {year} {2022})},\ \Eprint
  {http://arxiv.org/abs/2206.13208} {arXiv:2206.13208 [nucl-th]} \BibitemShut
  {NoStop}%
\bibitem [{\citenamefont {Gnech}\ \emph {et~al.}(2024)\citenamefont {Gnech},
  \citenamefont {Fore}, \citenamefont {Tropiano},\ and\ \citenamefont
  {Lovato}}]{Gnech24}%
  \BibitemOpen
  \bibfield  {author} {\bibinfo {author} {\bibfnamefont {Alex}\ \bibnamefont
  {Gnech}}, \bibinfo {author} {\bibfnamefont {Bryce}\ \bibnamefont {Fore}},
  \bibinfo {author} {\bibfnamefont {Anthony~J.}\ \bibnamefont {Tropiano}}, \
  and\ \bibinfo {author} {\bibfnamefont {Alessandro}\ \bibnamefont {Lovato}},\
  }\bibfield  {title} {\enquote {\bibinfo {title} {Distilling the essential
  elements of nuclear binding via neural-network quantum states},}\ }\href
  {\doibase 10.1103/PhysRevLett.133.142501} {\bibfield  {journal} {\bibinfo
  {journal} {Phys. Rev. Lett.}\ }\textbf {\bibinfo {volume} {133}},\ \bibinfo
  {pages} {142501} (\bibinfo {year} {2024})}\BibitemShut {NoStop}%
\bibitem [{\citenamefont {Fore}\ \emph {et~al.}(2024)\citenamefont {Fore},
  \citenamefont {Kim}, \citenamefont {Hjorth-Jensen},\ and\ \citenamefont
  {Lovato}}]{Fore:2024exa}%
  \BibitemOpen
  \bibfield  {author} {\bibinfo {author} {\bibfnamefont {Bryce}\ \bibnamefont
  {Fore}}, \bibinfo {author} {\bibfnamefont {Jane}\ \bibnamefont {Kim}},
  \bibinfo {author} {\bibfnamefont {Morten}\ \bibnamefont {Hjorth-Jensen}}, \
  and\ \bibinfo {author} {\bibfnamefont {Alessandro}\ \bibnamefont {Lovato}},\
  }\bibfield  {title} {\enquote {\bibinfo {title} {{Investigating the crust of
  neutron stars with neural-network quantum states}},}\ }\href@noop {} {\
  (\bibinfo {year} {2024})},\ \Eprint {http://arxiv.org/abs/2407.21207}
  {arXiv:2407.21207 [nucl-th]} \BibitemShut {NoStop}%
\bibitem [{\citenamefont {Efros}\ \emph {et~al.}(2007)\citenamefont {Efros},
  \citenamefont {Leidemann}, \citenamefont {Orlandini},\ and\ \citenamefont
  {Barnea}}]{Efros:2007nq}%
  \BibitemOpen
  \bibfield  {author} {\bibinfo {author} {\bibfnamefont {V.~D.}\ \bibnamefont
  {Efros}}, \bibinfo {author} {\bibfnamefont {W.}~\bibnamefont {Leidemann}},
  \bibinfo {author} {\bibfnamefont {G.}~\bibnamefont {Orlandini}}, \ and\
  \bibinfo {author} {\bibfnamefont {N.}~\bibnamefont {Barnea}},\ }\bibfield
  {title} {\enquote {\bibinfo {title} {{The Lorentz Integral Transform (LIT)
  method and its applications to perturbation induced reactions}},}\ }\href
  {\doibase 10.1088/0954-3899/34/12/R02} {\bibfield  {journal} {\bibinfo
  {journal} {J. Phys. G}\ }\textbf {\bibinfo {volume} {34}},\ \bibinfo {pages}
  {R459--2456} (\bibinfo {year} {2007})},\ \Eprint
  {http://arxiv.org/abs/0708.2803} {arXiv:0708.2803 [nucl-th]} \BibitemShut
  {NoStop}%
\bibitem [{\citenamefont {Hendry}\ and\ \citenamefont
  {Feiguin}(2019)}]{Hendry:2019}%
  \BibitemOpen
  \bibfield  {author} {\bibinfo {author} {\bibfnamefont {Douglas}\ \bibnamefont
  {Hendry}}\ and\ \bibinfo {author} {\bibfnamefont {Adrian~E.}\ \bibnamefont
  {Feiguin}},\ }\bibfield  {title} {\enquote {\bibinfo {title} {Machine
  learning approach to dynamical properties of quantum many-body systems},}\
  }\href {\doibase 10.1103/PhysRevB.100.245123} {\bibfield  {journal} {\bibinfo
   {journal} {Phys. Rev. B}\ }\textbf {\bibinfo {volume} {100}},\ \bibinfo
  {pages} {245123} (\bibinfo {year} {2019})}\BibitemShut {NoStop}%
\bibitem [{\citenamefont {{K{\"u}hner}}\ and\ \citenamefont
  {{White}}(1999)}]{Kuhner:1999}%
  \BibitemOpen
  \bibfield  {author} {\bibinfo {author} {\bibfnamefont {Till~D.}\ \bibnamefont
  {{K{\"u}hner}}}\ and\ \bibinfo {author} {\bibfnamefont {Steven~R.}\
  \bibnamefont {{White}}},\ }\bibfield  {title} {\enquote {\bibinfo {title}
  {{Dynamical correlation functions using the density matrix renormalization
  group}},}\ }\href {\doibase 10.1103/PhysRevB.60.335} {\bibfield  {journal}
  {\bibinfo  {journal} {\prb}\ }\textbf {\bibinfo {volume} {60}},\ \bibinfo
  {pages} {335--343} (\bibinfo {year} {1999})},\ \Eprint
  {http://arxiv.org/abs/cond-mat/9812372} {arXiv:cond-mat/9812372 [cond-mat]}
  \BibitemShut {NoStop}%
\bibitem [{\citenamefont {Carlson}\ and\ \citenamefont
  {Schiavilla}(1992)}]{Carlson:1992ga}%
  \BibitemOpen
  \bibfield  {author} {\bibinfo {author} {\bibfnamefont {J.}~\bibnamefont
  {Carlson}}\ and\ \bibinfo {author} {\bibfnamefont {R.}~\bibnamefont
  {Schiavilla}},\ }\bibfield  {title} {\enquote {\bibinfo {title} {{Euclidean
  proton response in light nuclei}},}\ }\href {\doibase
  10.1103/PhysRevLett.68.3682} {\bibfield  {journal} {\bibinfo  {journal}
  {Phys. Rev. Lett.}\ }\textbf {\bibinfo {volume} {68}},\ \bibinfo {pages}
  {3682--3685} (\bibinfo {year} {1992})}\BibitemShut {NoStop}%
\bibitem [{\citenamefont {Carlson}\ and\ \citenamefont
  {Schiavilla}(1994)}]{Carlson:1994zz}%
  \BibitemOpen
  \bibfield  {author} {\bibinfo {author} {\bibfnamefont {J.}~\bibnamefont
  {Carlson}}\ and\ \bibinfo {author} {\bibfnamefont {R.}~\bibnamefont
  {Schiavilla}},\ }\bibfield  {title} {\enquote {\bibinfo {title} {{Inclusive
  electron scattering and pion degrees of freedom in light nuclei}},}\ }\href
  {\doibase 10.1103/PhysRevC.49.R2880} {\bibfield  {journal} {\bibinfo
  {journal} {Phys. Rev. C}\ }\textbf {\bibinfo {volume} {49}},\ \bibinfo
  {pages} {R2880--R2884} (\bibinfo {year} {1994})}\BibitemShut {NoStop}%
\bibitem [{\citenamefont {Kim}\ \emph {et~al.}(2024)\citenamefont {Kim},
  \citenamefont {Pescia}, \citenamefont {Fore}, \citenamefont {Nys},
  \citenamefont {Carleo}, \citenamefont {Gandolfi}, \citenamefont
  {Hjorth-Jensen},\ and\ \citenamefont {Lovato}}]{Kim:2023fwy}%
  \BibitemOpen
  \bibfield  {author} {\bibinfo {author} {\bibfnamefont {Jane}\ \bibnamefont
  {Kim}}, \bibinfo {author} {\bibfnamefont {Gabriel}\ \bibnamefont {Pescia}},
  \bibinfo {author} {\bibfnamefont {Bryce}\ \bibnamefont {Fore}}, \bibinfo
  {author} {\bibfnamefont {Jannes}\ \bibnamefont {Nys}}, \bibinfo {author}
  {\bibfnamefont {Giuseppe}\ \bibnamefont {Carleo}}, \bibinfo {author}
  {\bibfnamefont {Stefano}\ \bibnamefont {Gandolfi}}, \bibinfo {author}
  {\bibfnamefont {Morten}\ \bibnamefont {Hjorth-Jensen}}, \ and\ \bibinfo
  {author} {\bibfnamefont {Alessandro}\ \bibnamefont {Lovato}},\ }\bibfield
  {title} {\enquote {\bibinfo {title} {{Neural-network quantum states for
  ultra-cold Fermi gases}},}\ }\href {\doibase 10.1038/s42005-024-01613-w}
  {\bibfield  {journal} {\bibinfo  {journal} {Commun. Phys.}\ }\textbf
  {\bibinfo {volume} {7}},\ \bibinfo {pages} {148} (\bibinfo {year} {2024})},\
  \Eprint {http://arxiv.org/abs/2305.08831} {arXiv:2305.08831
  [cond-mat.quant-gas]} \BibitemShut {NoStop}%
\bibitem [{\citenamefont {Schiavilla}\ \emph {et~al.}(2021)\citenamefont
  {Schiavilla}, \citenamefont {Girlanda}, \citenamefont {Gnech}, \citenamefont
  {Kievsky}, \citenamefont {Lovato}, \citenamefont {Marcucci}, \citenamefont
  {Piarulli},\ and\ \citenamefont {Viviani}}]{Schiavilla:2021dun}%
  \BibitemOpen
  \bibfield  {author} {\bibinfo {author} {\bibfnamefont {R.}~\bibnamefont
  {Schiavilla}}, \bibinfo {author} {\bibfnamefont {L.}~\bibnamefont
  {Girlanda}}, \bibinfo {author} {\bibfnamefont {A.}~\bibnamefont {Gnech}},
  \bibinfo {author} {\bibfnamefont {A.}~\bibnamefont {Kievsky}}, \bibinfo
  {author} {\bibfnamefont {A.}~\bibnamefont {Lovato}}, \bibinfo {author}
  {\bibfnamefont {L.~E.}\ \bibnamefont {Marcucci}}, \bibinfo {author}
  {\bibfnamefont {M.}~\bibnamefont {Piarulli}}, \ and\ \bibinfo {author}
  {\bibfnamefont {M.}~\bibnamefont {Viviani}},\ }\bibfield  {title} {\enquote
  {\bibinfo {title} {{Two- and three-nucleon contact interactions and
  ground-state energies of light- and medium-mass nuclei}},}\ }\href {\doibase
  10.1103/PhysRevC.103.054003} {\bibfield  {journal} {\bibinfo  {journal}
  {Phys. Rev. C}\ }\textbf {\bibinfo {volume} {103}},\ \bibinfo {pages}
  {054003} (\bibinfo {year} {2021})},\ \Eprint
  {http://arxiv.org/abs/2102.02327} {arXiv:2102.02327 [nucl-th]} \BibitemShut
  {NoStop}%
\bibitem [{\citenamefont {Sinibaldi}\ \emph {et~al.}(2023)\citenamefont
  {Sinibaldi}, \citenamefont {Giuliani}, \citenamefont {Carleo},\ and\
  \citenamefont {Vicentini}}]{Sinibaldi:2023yoq}%
  \BibitemOpen
  \bibfield  {author} {\bibinfo {author} {\bibfnamefont {Alessandro}\
  \bibnamefont {Sinibaldi}}, \bibinfo {author} {\bibfnamefont {Clemens}\
  \bibnamefont {Giuliani}}, \bibinfo {author} {\bibfnamefont {Giuseppe}\
  \bibnamefont {Carleo}}, \ and\ \bibinfo {author} {\bibfnamefont {Filippo}\
  \bibnamefont {Vicentini}},\ }\bibfield  {title} {\enquote {\bibinfo {title}
  {{Unbiasing time-dependent Variational Monte Carlo by projected quantum
  evolution}},}\ }\href {\doibase 10.22331/q-2023-10-10-1131} {\bibfield
  {journal} {\bibinfo  {journal} {Quantum}\ }\textbf {\bibinfo {volume} {7}},\
  \bibinfo {pages} {1131} (\bibinfo {year} {2023})},\ \Eprint
  {http://arxiv.org/abs/2305.14294} {arXiv:2305.14294 [quant-ph]} \BibitemShut
  {NoStop}%
\bibitem [{\citenamefont {Sorella}(2005)}]{Sorella:2005}%
  \BibitemOpen
  \bibfield  {author} {\bibinfo {author} {\bibfnamefont {Sandro}\ \bibnamefont
  {Sorella}},\ }\bibfield  {title} {\enquote {\bibinfo {title} {{Wave function
  optimization in the variational Monte Carlo method}},}\ }\href {\doibase
  10.1103/PhysRevB.71.241103} {\bibfield  {journal} {\bibinfo  {journal} {Phys.
  Rev. B}\ }\textbf {\bibinfo {volume} {71}},\ \bibinfo {pages} {241103}
  (\bibinfo {year} {2005})}\BibitemShut {NoStop}%
\bibitem [{\citenamefont {Gravina}\ \emph {et~al.}(2024)\citenamefont
  {Gravina}, \citenamefont {Savona},\ and\ \citenamefont
  {Vicentini}}]{Gravina:2024elh}%
  \BibitemOpen
  \bibfield  {author} {\bibinfo {author} {\bibfnamefont {Luca}\ \bibnamefont
  {Gravina}}, \bibinfo {author} {\bibfnamefont {Vincenzo}\ \bibnamefont
  {Savona}}, \ and\ \bibinfo {author} {\bibfnamefont {Filippo}\ \bibnamefont
  {Vicentini}},\ }\bibfield  {title} {\enquote {\bibinfo {title} {{Neural
  Projected Quantum Dynamics: a systematic study}},}\ }\href@noop {} {\
  (\bibinfo {year} {2024})},\ \Eprint {http://arxiv.org/abs/2410.10720}
  {arXiv:2410.10720 [quant-ph]} \BibitemShut {NoStop}%
\bibitem [{\citenamefont {Jarrell}\ and\ \citenamefont
  {Gubernatis}(1996)}]{Jarrell:1996rrw}%
  \BibitemOpen
  \bibfield  {author} {\bibinfo {author} {\bibfnamefont {Mark}\ \bibnamefont
  {Jarrell}}\ and\ \bibinfo {author} {\bibfnamefont {J.~E.}\ \bibnamefont
  {Gubernatis}},\ }\bibfield  {title} {\enquote {\bibinfo {title} {{Bayesian
  inference and the analytic continuation of imaginary-time quantum Monte Carlo
  data}},}\ }\href {\doibase 10.1016/0370-1573(95)00074-7} {\bibfield
  {journal} {\bibinfo  {journal} {Phys. Rept.}\ }\textbf {\bibinfo {volume}
  {269}},\ \bibinfo {pages} {133--195} (\bibinfo {year} {1996})}\BibitemShut
  {NoStop}%
\bibitem [{\citenamefont {Arenhovel}\ and\ \citenamefont
  {Sanzone}(1991)}]{Arenhovel:1990yg}%
  \BibitemOpen
  \bibfield  {author} {\bibinfo {author} {\bibfnamefont {H}~\bibnamefont
  {Arenhovel}}\ and\ \bibinfo {author} {\bibfnamefont {M.}~\bibnamefont
  {Sanzone}},\ }\bibfield  {title} {\enquote {\bibinfo {title}
  {{Photodisintegration of the deuteron: A Review of theory and experiment}},}\
  }\href@noop {} {\bibfield  {journal} {\bibinfo  {journal} {Few Body Syst.
  Suppl.}\ }\textbf {\bibinfo {volume} {3}},\ \bibinfo {pages} {1--183}
  (\bibinfo {year} {1991})}\BibitemShut {NoStop}%
\bibitem [{\citenamefont {Siegert}(1937)}]{Siegert:1937yt}%
  \BibitemOpen
  \bibfield  {author} {\bibinfo {author} {\bibfnamefont {A.~J.~F.}\
  \bibnamefont {Siegert}},\ }\bibfield  {title} {\enquote {\bibinfo {title}
  {{Note on the interaction between nuclei and electromagnetic radiation}},}\
  }\href {\doibase 10.1103/PhysRev.52.787} {\bibfield  {journal} {\bibinfo
  {journal} {Phys. Rev.}\ }\textbf {\bibinfo {volume} {52}},\ \bibinfo {pages}
  {787--789} (\bibinfo {year} {1937})}\BibitemShut {NoStop}%
\bibitem [{\citenamefont {Barnea}\ \emph {et~al.}(2000)\citenamefont {Barnea},
  \citenamefont {Leidemann},\ and\ \citenamefont {Orlandini}}]{Barnea:1999be}%
  \BibitemOpen
  \bibfield  {author} {\bibinfo {author} {\bibfnamefont {Nir}\ \bibnamefont
  {Barnea}}, \bibinfo {author} {\bibfnamefont {Winfried}\ \bibnamefont
  {Leidemann}}, \ and\ \bibinfo {author} {\bibfnamefont {Giuseppina}\
  \bibnamefont {Orlandini}},\ }\bibfield  {title} {\enquote {\bibinfo {title}
  {{State dependent effective interaction for the hyperspherical formalism}},}\
  }\href {\doibase 10.1103/PhysRevC.61.054001} {\bibfield  {journal} {\bibinfo
  {journal} {Phys. Rev. C}\ }\textbf {\bibinfo {volume} {61}},\ \bibinfo
  {pages} {054001} (\bibinfo {year} {2000})},\ \Eprint
  {http://arxiv.org/abs/nucl-th/9910062} {arXiv:nucl-th/9910062} \BibitemShut
  {NoStop}%
\bibitem [{\citenamefont {Bacca}\ \emph {et~al.}(2014)\citenamefont {Bacca},
  \citenamefont {Barnea}, \citenamefont {Hagen}, \citenamefont {Miorelli},
  \citenamefont {Orlandini},\ and\ \citenamefont {Papenbrock}}]{Bacca:2014rta}%
  \BibitemOpen
  \bibfield  {author} {\bibinfo {author} {\bibfnamefont {S.}~\bibnamefont
  {Bacca}}, \bibinfo {author} {\bibfnamefont {N.}~\bibnamefont {Barnea}},
  \bibinfo {author} {\bibfnamefont {G.}~\bibnamefont {Hagen}}, \bibinfo
  {author} {\bibfnamefont {M.}~\bibnamefont {Miorelli}}, \bibinfo {author}
  {\bibfnamefont {G.}~\bibnamefont {Orlandini}}, \ and\ \bibinfo {author}
  {\bibfnamefont {T.}~\bibnamefont {Papenbrock}},\ }\bibfield  {title}
  {\enquote {\bibinfo {title} {{Giant and pigmy dipole resonances in $^4$He,
  $^{16,22}$O, and $^{40}$Ca from chiral nucleon-nucleon interactions}},}\
  }\href {\doibase 10.1103/PhysRevC.90.064619} {\bibfield  {journal} {\bibinfo
  {journal} {Phys. Rev. C}\ }\textbf {\bibinfo {volume} {90}},\ \bibinfo
  {pages} {064619} (\bibinfo {year} {2014})},\ \Eprint
  {http://arxiv.org/abs/1410.2258} {arXiv:1410.2258 [nucl-th]} \BibitemShut
  {NoStop}%
\bibitem [{\citenamefont {Skilling}(1989)}]{Skilling1989}%
  \BibitemOpen
  \bibfield  {author} {\bibinfo {author} {\bibfnamefont {John}\ \bibnamefont
  {Skilling}},\ }\enquote {\bibinfo {title} {Classic maximum entropy},}\ in\
  \href {\doibase 10.1007/978-94-015-7860-8_3} {\emph {\bibinfo {booktitle}
  {Maximum Entropy and Bayesian Methods: Cambridge, England, 1988}}},\ \bibinfo
  {editor} {edited by\ \bibinfo {editor} {\bibfnamefont {J.}~\bibnamefont
  {Skilling}}}\ (\bibinfo  {publisher} {Springer Netherlands},\ \bibinfo
  {address} {Dordrecht},\ \bibinfo {year} {1989})\ pp.\ \bibinfo {pages}
  {45--52}\BibitemShut {NoStop}%
\end{thebibliography}%

\clearpage

\onecolumngrid

\begin{center}
    {\Large \textbf{Supplemental Material}}
\end{center}

\setcounter{equation}{0}
\setcounter{figure}{0}
\setcounter{table}{0}
\setcounter{section}{0}

\section{Fidelity optimization}
We solve the LIT equation by maximizing the quantum fidelity between the states \( |\Psi\rangle \equiv (\hat{H}-z)|\Psi_L\rangle \) and \( |\Phi\rangle \equiv \hat{O} | \Psi_0 \rangle \)  
\begin{equation}
\mathcal{F}(\Psi, \Phi) = \frac{\langle \Psi | \Phi \rangle \langle \Phi | \Psi \rangle }{ \langle \Psi | \Psi \rangle \langle \Phi | \Phi \rangle}.
\end{equation}

We compute this quantity stochastically using the double Monte Carlo estimator~\cite{Gravina:2024elh}
\begin{align}
\mathcal{F}(\Psi, \Phi) &= \mathbb{E}_{X\sim \pi_\Psi} [H_{\rm loc}(X)], \nonumber \\
H_{\rm loc}(X) &= \frac{\Phi(X)}{\Psi(X)} \mathbb{E}_{Y\sim \pi_\Phi}\left[  \frac{\Psi(Y)}{\Phi(Y)}\right],
\end{align}
where we introduce the probability distributions  
\begin{equation}
 \pi_\Psi(X) = \frac{|\Psi(X)|^2}{\langle \Psi | \Psi \rangle}, \qquad  
 \pi_\Phi(X) = \frac{|\Phi(X)|^2}{\langle \Phi | \Phi \rangle}\, .
\end{equation}

Recently, control-variate versions of this estimator have been derived. However, their impact on variance reduction is limited, as the key factor affecting optimization is the gradient estimator rather than the fidelity estimator itself. In this work, we employ the following gradient estimator:  
\begin{equation}
\nabla_{\theta}\mathcal{F} =  \mathbb{E}_{X\sim \pi_\Psi}  [2 \operatorname{Re}\{\Delta J(X) H_{\rm loc}^*(X)\}],
\end{equation}
a full derivation of which can be found in Appendix E3 of Ref.~\cite{Gravina:2024elh}. The centered Jacobian is given by  
\begin{equation}
\Delta J(X) = \nabla_{\theta}\log \Psi(X) - \mathbb{E}_{X\sim \pi_\Psi}[\nabla_{\theta}\log \Psi(X)]\,,
\end{equation}
ensuring that the above fidelity estimate is itself a covariance. This property significantly reduces statistical fluctuations, improving the convergence of the optimization.  

Note that the quantum Fisher information matrix of Eq.~\ref{eq:fisher} can be readily estimated as
\begin{equation}
S = \mathbb{E}_{X\sim \pi_\Psi}[\Delta J^\dagger(X) \Delta J(X)]\, .
\end{equation}

Finally, although computing the above estimators in principle requires sampling from both \( \pi_\Psi \) and \( \pi_\Phi \), we leverage importance sampling to sample from \( \pi_\Phi \) only. The latter approach is particularly efficient in this context, as \( \Phi(X) \) does not depend on \( \omega_0 \). Consequently, a large number of samples can be generated at the beginning of the calculation and then reused for all values of \( \omega_0 \).  

We find it convenient to begin the calculation of the LIT by first considering a very large and negative value of \( \omega_0 = -100 \) MeV, as the solution \( |\Psi_L\rangle \) in this region is proportional to the source \( \hat{O}|\Psi_0\rangle \). To accelerate the minimization of the fidelity, we do not start from randomly initialized parameters but instead recycle those associated with the ground-state wave function \( \Psi_0(X) \). Owing to the nature of the dipole transition operator, the parity of \( |\Psi_L\rangle \) is chosen to be opposite to that of the ground state—both \( ^2\mathrm{H} \) and \( ^4\mathrm{He} \) have positive-parity ground states, and hence \( |\Psi_L\rangle \) has negative parity. Once the fidelity for the smallest value of \( \Gamma \) has reached a plateau, we employ transfer-learning techniques, using the best parameters obtained as the starting point for the next \( \omega_0 \). This procedure is repeated until we reach the maximum value of \( \omega_0 = 200 \) MeV, which is relevant for the dipole response.

\section{Detailed derivation of the error bounds}
Presented here is the first analytical bound for the errors of
a LIT method. This bound lemma is universal in the sense that it doesn't depend on any specific LIT solving method and can be applied to all of them. Without loss of generality, we can write the optimized function as $|\bar{\Psi}_L\rangle=\mathcal{N}|\Psi_L\rangle+\mathcal{M}|\Psi_\perp\rangle$ with $(H-z)|\Psi_L\rangle=|\Phi\rangle$, 
$(H-z)|\Psi_\perp\rangle\perp|\Phi\rangle$, $\operatorname{Im}(\mathcal{M})=0$ and $||\Psi_\perp\rangle|=1$. We remind the notation $|\bar{\Psi}\rangle =(H-z)|\bar{\Psi}_L\rangle $.
The constants $\mathcal{N} $ and $\mathcal{M}$ can be computed via:
\begin{equation}
\mathcal{N}=\frac{\langle \Phi|\Psi\rangle}{\langle \Phi|\Phi\rangle},\mathcal{M}^{2}=\frac{\left|(1-P_{\Phi})|\Psi\rangle\right|^{2}}{\langle\Psi_\perp|\tilde{H}^{\dagger}\tilde{H}|\Psi_\perp\rangle}\label{eq:alpha-1}
\end{equation}
where we introduced $\tilde{H} = H-z$ and $P_{\Phi}=\frac{|\Phi\rangle\langle \Phi|}{\langle \Phi|\Phi\rangle}$ is the projection operator over $|\Phi\rangle$. As discussed in the main text, the LIT is computed in a numerically stable fashion by 
\begin{equation}
\mathcal{L}=\langle\Psi_L|\Psi_L\rangle=\langle \Phi|\frac{1}{H-z^{*}}\frac{1}{H-z}|\Phi\rangle = \frac{-2}{z^{*}-z}\operatorname{Im}\left(\langle \Phi|\frac{1}{H-z}|\Phi\rangle\right) = - \frac{1}{\Gamma}\operatorname{Im}\left(\langle \Phi|\Psi_L\rangle\right).
\label{eq:LIT_overlap}
\end{equation}

Expressing $\Psi_L$ in terms of $\bar{\Psi}_L$ and $\Psi_\perp\rangle$ yields
\[
\mathcal{L}=-\frac{1}{\Gamma}\operatorname{Im}\left(\frac{1}{\mathcal{N}}\langle \Phi|\bar{\Psi}_L\rangle\right)+\frac{1}{\Gamma}\operatorname{Im}\left(\frac{\mathcal{M}}{\mathcal{N}}\langle \Phi|\Psi_\perp\rangle\right).
\]
We will denote the first term $\mathcal{L}_{calc}$ and the second $\Delta$. We can now provide an error bound for computed LIT by  bounding $\Delta$. To this end, we note that:
\[
\left|\Delta\right|\leq\frac{\left|\langle \Phi|\Psi_\perp\rangle\right|}{\Gamma\left|\tilde{H}|\Psi_\perp\rangle\right|}\frac{|\Phi|^{2}}{\left|\langle \Phi|\Psi\rangle\right|}\left|(1-P_{\Phi})|\Psi\rangle\right|.
\]
We can simplify the above expression using the definition of $\mathcal{F}$ and noting that $\left|(1-P_{\Phi})|\Psi\rangle\right|=|\Psi|\sqrt{1-\mathcal{F}}$, so that
 \[
\left|\Delta\right|\leq\frac{\left|\langle \Phi|\Psi_\perp\rangle\right|}{\left|\tilde{H}|\Psi_\perp\rangle\right|}\frac{|\Phi|}{\Gamma}\sqrt{\frac{1-\mathcal{F}}{\mathcal{F}}}.
\]
Placing an upper limit on the error is equivalent to seeking the largest value  of $\left|\langle \Phi|\Psi_\perp\rangle\right|/\left|\tilde{H}|\Psi_\perp\rangle\right|$. First, we can take an optimization problem similar to the LIT equation. Using this understanding we write
\begin{equation*}
\frac{\left|\langle \Phi|\Psi_\perp\rangle\right|}{\left|\tilde{H}|\Psi_\perp\rangle\right|}=
\frac{\left|\langle \Phi|\tilde{H}^{-1}\tilde{H}|\Psi_\perp\rangle\right|}{\left|\tilde{H}|\Psi_\perp\rangle\right|}=
\frac{\left|\langle \Phi|\tilde{H}^{-1}\left(1-P_\Phi\right)\tilde{H}|\Psi_\perp\rangle\right|}{\left|\tilde{H}|\Psi_\perp\rangle\right|}\leq
\left|\langle \Phi|\tilde{H}^{-1}\left(1-P_\Phi\right)\right|=
\left|\left(1-P_{\Phi}\right)|\Psi_{L}\rangle\right| \approx
\frac{1}{|\mathcal{N}|}\left|\left(1-P_{\Phi}\right)|\bar{\Psi}_L\rangle\right|\, ,
\end{equation*}
thus
\begin{equation}
\left|\Delta\right|\lesssim \frac{1}{|\mathcal{N}|}\left|\left(1-P_{\Phi}\right)|\Psi_L\rangle\right| \frac{|\Phi|}{\Gamma}\sqrt{\frac{1-\mathcal{F}}{\mathcal{F}}}.
\label{eq:bound_1}
\end{equation}
Another error bound can be obtained by noting that
\begin{equation*}
\langle \Phi|H-\sigma-i\Gamma|\Psi_\perp\rangle=\langle \Phi|\tilde{H}|\Psi_\perp\rangle=0\Longrightarrow\langle \Phi|\Psi_\perp\rangle=\frac{\langle \Phi|H|\Psi_\perp\rangle}{\left(\sigma+i\Gamma\right)}
\end{equation*}
which implies that
\[\frac{\left|\langle\Phi|\Psi_\perp\rangle\right|}{\left|\tilde{H}|\Psi_\perp\rangle\right|}=
\frac{1}{\left|\tilde{H}|\Psi_\perp\rangle\right|}\frac{|\langle\Phi|H|\Psi_\perp\rangle|}{\left|\sigma+i\Gamma\right|}=
\frac{1}{\left|\tilde{H}|\Psi_\perp\rangle\right|}\frac{1}{\left|\sigma+i\Gamma\right|}\left|\langle\Phi|\frac{H}{\tilde{H}}\tilde{H}|\Psi_\perp\rangle\right|=
\frac{1}{\left|\tilde{H}|\Psi_\perp\rangle\right|}\frac{1}{\left|\sigma+i\Gamma\right|}\left|\langle\Phi|\frac{H}{\tilde{H}}\left(1-P_{\Phi}\right)\tilde{H}|\Psi_\perp\rangle\right|\]
\[\leq\frac{1}{\sqrt{\sigma^{2}+\Gamma^{2}}}\left|\left(1-P_{\Phi}\right)\frac{H}{\tilde{H}}|\Phi\rangle\right|=
\frac{1}{\sqrt{\sigma^{2}+\Gamma^{2}}}\left|\left(1-P_{\Phi}\right)H|\Psi_L\rangle\right|\approx\frac{1}{|\mathcal{N}|}\frac{1}{\sqrt{\sigma^{2}+\Gamma^{2}}}\left|\left(1-P_{\Phi}\right)H|\bar{\Psi}_L\rangle\right|\]
Plugging this back gives
\[\left|\Delta\right|\lesssim\frac{|\Phi|}{\Gamma|\mathcal{N}|}\frac{1}{\sqrt{\sigma^{2}+\Gamma^{2}}}\left|\left(1-P_{\Phi}\right)H|\bar{\Psi}_L\rangle\right|\sqrt{\frac{1-\mathcal{F}}{\mathcal{F}}}.\]

We can consider both bounds together by 
\begin{equation}
\Delta\mathcal{L}(\omega_{0},\Gamma)\leq \min\left(\left|\left(1-P_{\Phi}\right)|\bar{\Psi}_L\rangle\right|,\left|\left(1-P_{\Phi}\right)\frac{H}{\sqrt{\sigma^{2}+\Gamma^{2}}}|\bar{\Psi}_L\rangle\right|\right)\frac{|\Phi|}{\Gamma |\mathcal{N}|}\sqrt{\frac{1-\mathcal{F}}{\mathcal{F}}}.
\label{eq:LIT_bound_full}
\end{equation}
Note that for a well-optimized wavefunction $1-\mathcal{F}\to0$ and so $|\Delta|\to0$. This gives us also some understanding of problem difficulty as a function of the parameters: lower $\Gamma$ and/or lower $\mathcal{L}$ require better convergence in $\sqrt{1-\mathcal{F}}$ for a given relative accuracy.

\section{Inversion of the LIT}
The inversion of the LIT proceeds in two steps. The first step is a regularized variant of the standard inversion procedure introduced in Ref.~\cite{Efros:2007nq}, which expands the response function as
\begin{equation}
R(\omega) = \sum_{n=1}^{N} c_{n}\,\chi_{n}(\omega^\prime,\alpha_1,\alpha_2)\,,
\label{eq:res_basis}
\end{equation}
where $\omega^\prime = \omega - \omega_{\rm th}$ and $\omega_{\rm th}$ denotes the threshold energy for breakup into the continuum. The basis functions, which depend nonlinearly on $\alpha_1$ and $\alpha_2$, are given by
\begin{equation}
\chi_{1}(\omega,\alpha_1,\alpha_2)=\omega^{\alpha_{1}}e^{-\alpha_{2}\omega}\quad,\quad 
\chi_{n\geq 2} (\omega,\alpha_1,\alpha_2)= \omega^{\alpha_{1}}\exp\left(-\frac{\alpha_{2}\omega}{n}\right)\,.
\end{equation}
The LIT of the response function in Eq.~\eqref{eq:res_basis} reads
\begin{equation}
\tilde{\mathcal{L}}(\omega_0)=\sum_{n=1}^{M} c_{n}\,\tilde{\chi}_{n}\left(\omega_0, \alpha_1, \alpha_2\right)\,,
\end{equation}
where the LIT of each basis function is defined as
\begin{equation}
\tilde{\chi}_{n}\left(\omega_0, \alpha_1, \alpha_2\right)=\int_{0}^{\infty} d\omega^\prime\,\frac{\chi_{n}\left(\omega^\prime, \alpha_1, \alpha_2\right)}{(\omega^\prime-\omega_0)^{2}+\Gamma^{2}}\,.
\end{equation}
We determine the optimal parameters $c_N$ by minimizing a chi-squared loss function on a discretized grid of $N_{\omega_0}$ points in $\omega_0$ where the LIT is computed,
\begin{equation}
\chi^2=\sum_{i=1}^{N_{\omega_0}} \frac{(\tilde{\mathcal{L}}_i-\mathcal{L}_i)^2}{\sigma^2_i}\,.
\end{equation}
For simplicity, in this first inversion method we assume unit errors, $\sigma^2_i=1$. 
To stabilize the inversion, in analogy with methods commonly employed in deep neural networks, we add an $L^{2}$ regularization term on the linear parameters,
\begin{equation}
I=I_{0}+\gamma\sum_{n=1}^{N} c_{n}^{2}\,.
\end{equation}
We find that choosing values of the hyperparameter $\gamma$ in the range $10^{-7}$--$10^{-2}$ stabilizes the inversion by preventing the parameters from becoming exceedingly large and compensating each other. Crucially, it also reduces the dependence of the final result on the number $N$ of basis functions used.

The second step is an improved version of the celebrated maximum-entropy approach~\cite{Jarrell:1996rrw}, which seeks the response function $R$ that maximizes the conditional probability given the LIT data $\mathcal{L}$:
\begin{equation}
P(R\, | \mathcal{L}) = P(\mathcal{L} | R) \frac{P(R)}{P(\mathcal{L})}\,,
\end{equation}
where $P(\mathcal{L})$ is a normalization constant—since we work with one set of LIT data at a time, it can be safely ignored. The likelihood function is
\begin{equation}
P(\mathcal{L} | R) = \frac{e^{-\chi^2 / 2}}{Z_L}\,,
\end{equation}
When evaluating $\chi^2$, we account for both statistical and systematic errors, added in quadrature:
\begin{equation}
\sigma_i^2 = \sqrt{\sigma_{i\, \text{stat}}^2 + \sigma_{i\, \text{sys}}^2}\,,
\end{equation}
where the systematic error is given by Eq.~\eqref{eq:error} and the statical error is computed from evaluating stochastically the overlap as in Eq.~\eqref{eq:lit_xilin}. The normalization factor $Z_L$ can be computed analytically, as the integral is Gaussian:
\begin{equation}
Z_L = \int \prod_{i=1}^{N_{\omega_0}}d\mathcal{L}_i \, e^{-\chi^2 / 2} = (2\pi)^{N/2} \prod_{i=1}^{N_{\omega_0}} \sigma_i\,.
\end{equation}
Skilling~\cite{Skilling1989} argues that the prior probability $P(R)$ for an unnormalized probability density is proportional to $\exp(\alpha S)$, where $S$ is the entropy defined relative to some positive-definite function $m(\omega)$:
\begin{equation}
S = \int d\omega \left[ R(\omega) - m(\omega) - R(\omega) \log\left(\frac{R(\omega)}{m(\omega)}\right) \right] \simeq \sum_{i=1}^{N_\omega} \left[ R_i - m_i - R_i  \log\left(\frac{R_i}{m_i}\right) \right] \Delta\omega \,,
\end{equation}
where we defined $R_i = R(\omega_i)$ and $m_i = m(\omega_i)$. This expression reduces to the conventional entropy definition if both $R$ and $m$ are normalized to unity. Hence, the prior is conditional on both $m$ and $\alpha$, and can be written formally as
\begin{equation}
P(R\, | m, \alpha) = \frac{e^{\alpha S}}{Z_S} \simeq \frac{1}{Z_S}\exp \left\{\alpha \sum_{i=1}^{N_\omega} \left[ R_i - m_i - R_i  \log\left(\frac{R_i}{m_i}\right) \right]\right\}\,,
\end{equation}
which is maximized when $S = 0$, i.e., when $R = m$. The $\alpha$-dependent normalization constant $Z_S$ can be estimated by approximating the exponential with a Gaussian centered at $R_i=m_i$, yielding~\cite{Jarrell:1996rrw}
\begin{equation}
Z_S =   \int \prod_{i=1}^{N_\omega} \frac{d R_i}{\sqrt{R_i}}\exp \left\{\alpha\left[ R_i - m_i - R_i  \log\left(\frac{R_i}{m_i}\right) \right]\right\}\simeq \left(\frac{2 \pi}{\alpha}\right)^{N_\omega / 2}\, .
\end{equation}

Using Bayes' theorem, the posterior distribution of $R$ given the LIT data, the model, and $\alpha$ is
\begin{equation}
P(R | \mathcal{L}, m, \alpha)= \frac{e^{\alpha S -\chi^2 / 2}}{Z_S Z_L}\,.
\end{equation}
Existing maximum-entropy approaches rely on maximizing the above probability, which corresponds to minimizing $Q = \alpha S -\chi^2 / 2$. Clearly, the reconstructed spectrum $\tilde{R}_\alpha$ depends on the choice of $\alpha$. Specifically, {\it historic} maximum entropy picks the $\alpha$ that yields $\chi^2=1$ at the minimum of $Q$. On the other hand, the {\it classic} maximum entropy picks the value of $\alpha$ that maximizes the posterior probability
\begin{equation}
P(\alpha | \mathcal{L}, m ) = \int \prod_{i=1}^{N_\omega} \frac{d R_i}{\sqrt{R_i}} P(R, \alpha| \mathcal{L}, m ) 
= \int \prod_{i=1}^{N_\omega} \frac{d R_i}{\sqrt{R_i}} P(R | \mathcal{L}, m, \alpha)P(\alpha) 
= \int \prod_{i=1}^{N_\omega} \frac{d R_i}{\sqrt{R_i}} \frac{e^{\alpha S -\chi^2 / 2}}{Z_S Z_L} P(\alpha)\,.
\end{equation}
where $P(\alpha)$ is the prior probability of $\alpha$. Jeffreys argued that since $\chi^2/2$ and $S$ have different units, $\alpha$ is a scale factor and should be assigned the simplest scale-invariant form $P(\alpha) = 1 / \alpha$. Finally, {\it Bryan's} maximum-entropy approach averages $\tilde{R}_\alpha$ over this posterior. 

Here, as a significant departure from previous maximum entropy realizations, we use Monte Carlo techniques to directly sample the joint posterior distribution 
\begin{equation}
P(R, \alpha | \mathcal{L}, m)= \frac{e^{\alpha S -\chi^2 / 2}}{Z_S Z_L}P(\alpha)\,,
\end{equation}
Specifically, we parametrize $R_i = m_i e^u_i$ and use a Metropolis-Hastings algorithm to sample the set $\{u_i\}$. Owing to the high degree of correlation among adjacent $R_i$, in the Metropolis walk we both update $u_i$ by sampling from a Gaussian distribution and swap $u_i$ with $u_{i\pm 1}$. We find that this approach notably improves the ergodicity of the Markov chain.

The reconstructed response function is given by the mean spectrum 
\begin{equation}
\bar{R}(\omega) = \frac{1}{N_{\rm samples}} \sum_{R \sim P(R, \alpha | \mathcal{L}, m)} R(\omega)\,.
\end{equation}
In addition to its mean value, this procedure allows us to directly estimate the uncertainties of the reconstructed spectrum. As noted in Ref.~\cite{Jarrell:1996rrw}, it is not possible to assign error bars to each point in $\omega$, because errors at different points are strongly correlated. On the other hand, it is possible to assign error bars to integrated functions of
the spectral density:
\begin{equation}
H = \int d \omega \, R(\omega) \, h(\omega)\,,
\end{equation}
where $h(\omega)$ is an arbitrary function, which we choose to be a rectangular window function of width $1\,\mathrm{MeV}$.

\begin{figure}[!htb]
    \includegraphics[width=0.5\linewidth]{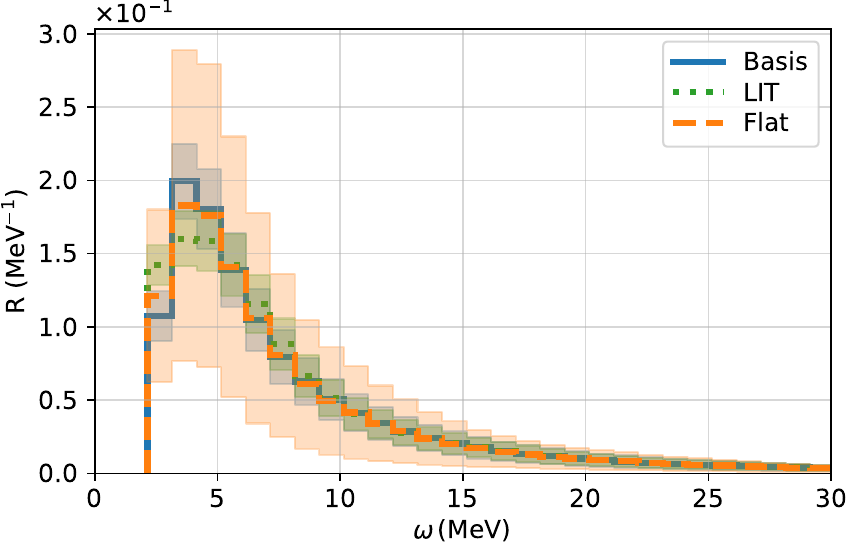}
    \caption{Dipole response function of $^2$H reconstructed from the corresponding LIT using the Maximum Entropy approach with three different prior models: basis function (solid blue), LIT (dotted green), and flat (dashed orange). \label{fig:h2_lit_prior}}
\end{figure}

The results displayed in the main body of the article use, as the default model, the response function expanded in a set of basis functions, as described in step one. However, we also tested how the reconstructed response depends on the choice of $m$ by considering (i) a flat prior between $\omega_{\rm th}$ and $\omega_{\max} = 200\,\mathrm{MeV}$, and (ii) the LIT itself, $m(\omega)=\mathcal{L}(\omega)$ for $\omega>\omega_{\rm th}$ and zero otherwise. For illustration, in Figure~\ref{fig:h2_lit_prior}, we display the reconstructed responses using the basis-function expansion, the flat model, and the LIT as default models. All reconstructions agree within the error bars. Consistent with Ref.~\cite{Jarrell:1996rrw}, the uncertainties associated with the least informative flat model are larger than those obtained when using either the LIT or the basis-function expansion as the default model.

\begin{figure}[!htb]
    \includegraphics[width=0.5\linewidth]{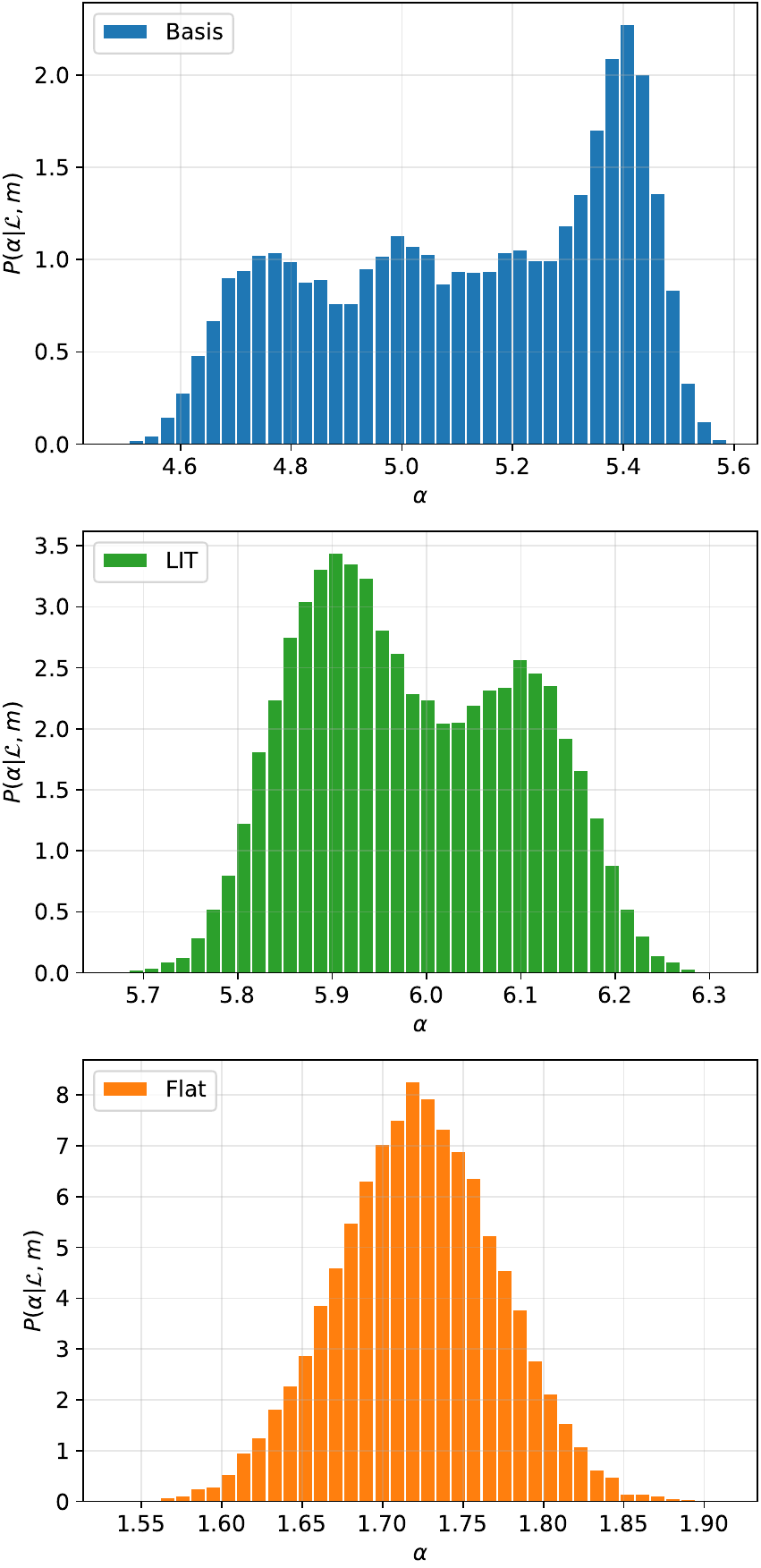}
    \caption{Posterior distribution of $\alpha$ corresponding to three different prior models: basis function (upper panel), LIT (middle panel), and flat (lower panel). \label{fig:h2_lit_alpha}}
\end{figure}

Using the samples drawn from $P(R, \alpha \mid \mathcal{L}, m)$, we can marginalize over $R$ to obtain the posterior distribution of $\alpha$, $P(\alpha \mid \mathcal{L}, m)$. The latter is shown in Figure~\ref{fig:h2_lit_alpha} for the three prior choices discussed above. As expected, for more informative models, $P(\alpha \mid \mathcal{L}, m)$ shifts toward larger values of $\alpha$, since minimizing $\chi^2$ does not drive the responses far from the prior. Additionally, we note that --- except for the flat prior case --- all posterior distributions $P(\alpha \mid \mathcal{L}, m)$ are neither Gaussian nor symmetric. Therefore, the classic maximum entropy approach, which entails fixing $\alpha$ to the maximum of the distribution, would not yield an accurate representation.

\end{document}